\documentclass[aps,pra,superscriptaddress,twocolumn,longbibliography]{revtex4-1}
\usepackage{units}
\usepackage{amsmath}
\usepackage{amsthm}
\usepackage{amssymb}
\usepackage{graphicx}
\usepackage{color}
\usepackage{xcolor}
\usepackage{bbold}

\usepackage[colorlinks=true, urlcolor=blue, citecolor=blue,linkcolor=blue,citebordercolor={1 0 0},linkbordercolor={0 0 1}]{hyperref}

\usepackage[linesnumbered, ruled,vlined]{algorithm2e}

\graphicspath{{./images/},{./imagesAppendix/}}

\renewcommand{\eqref}[1]{Eq.~(\ref{#1})} % Reference to equation

\theoremstyle{plain}

\theoremstyle{plain}

\ifx\proof\undefined
\newenvironment{proof}[1][\protect\proofname]{\par
	\normalfont\topsep6\p@\@plus6\p@\relax
	\trivlist
	\itemindent\parindent
	\item[\hskip\labelsep\scshape #1]\ignorespaces
}{%
	\endtrivlist\@endpefalse
}
\providecommand{\proofname}{Proof}
\fi
\theoremstyle{plain}

\theoremstyle{remark}

\newcommand{\bra}[1]{\langle #1|}
\newcommand{\ket}[1]{|#1 \rangle}
\makeatother
\newcommand{\braket}[2]{\langle #1 \vert #2 \rangle}
\newcommand{\abs}[1]{\left|#1\right|}
\newcommand{\idg}[1]{{\bfseries #1)}}

\newcommand\numberthis{\addtocounter{equation}{1}\tag{\theequation}}
\providecommand{\factname}{Fact}
\providecommand{\theoremname}{Theorem}
\providecommand{\claimname}{Claim}
\providecommand{\lemmaname}{Lemma}
\providecommand{\definitionname}{Definition}

\definecolor{KB}{rgb}{0.4,0.3,0.9}

\definecolor{THc}{rgb}{0.9,0.3,0.2}

\newcommand{\revA}[1]{{#1}}
\newcommand{\revB}[1]{#1}
\newcommand{\revC}[1]{ #1}

\theoremstyle{definition}

\newcommand{\subfigimg}[3][,]{%
	\setbox1=\hbox{\includegraphics[#1]{#3}}% Store image in box
	\leavevmode\rlap{\usebox1}% Print image
	\rlap{\hspace*{2pt}\raisebox{\dimexpr\ht1-0.5\baselineskip}{{\bfseries \large\textsf{#2}}}}% Print label
	\phantom{\usebox1}% Insert appropriate spcing
}

\begin{document}

\title{Quantum machine learning of large datasets using randomized measurements}

\author{Tobias Haug}
\email{tobias.haug@u.nus.edu}
\affiliation{QOLS, Blackett Laboratory, Imperial College London SW7 2AZ, UK}

\author{Chris N. Self}

\affiliation{QOLS, Blackett Laboratory, Imperial College London SW7 2AZ, UK}
\author{M. S. Kim}
\affiliation{QOLS, Blackett Laboratory, Imperial College London SW7 2AZ, UK}
\begin{abstract}
Quantum computers promise to enhance machine learning for practical applications.
Quantum machine learning for real-world data has to handle extensive amounts of high-dimensional data. However, conventional methods for measuring quantum kernels are impractical for large datasets as they scale with the square of the dataset size.
Here, we measure quantum kernels using randomized measurements. The quantum computation time scales linearly with dataset size and quadratic for classical post-processing. While our method scales in general exponentially in qubit number, we gain a substantial speed-up when running on intermediate-sized quantum computers.
Further, we efficiently encode high-dimensional data into quantum computers with the number of features scaling linearly with the circuit depth. 
The encoding is characterized by the quantum Fisher information metric and is related to the radial basis function kernel. Our approach is robust to noise via a cost-free error mitigation scheme.
We demonstrate the advantages of our methods for noisy quantum computers by classifying images with the IBM quantum computer. 
To achieve further speedups we distribute the quantum computational tasks between different quantum computers. 
Our method enables benchmarking of quantum machine learning algorithms with large datasets on currently available quantum computers.
\end{abstract}

\maketitle

\section{Introduction}
Quantum machine learning aims to use quantum computers to enhance the power of machine learning~\cite{biamonte2017quantum,schuld2018supervised}.
One possible route to quantum advantage in machine learning is the use of quantum embedding kernels~\cite{schuld2019quantum,schuld2021effect,lloyd2020quantum,li2021recent}, where quantum computers are used to encode data in ways that are difficult for classical machine learning methods~\cite{liu2020rigorous,huang2021power,huang2021provably}. 
Noisy intermediate scale quantum computers~\cite{preskill2018quantum,bharti2021noisy} may be capable of solving tasks difficult for classical computers~\cite{arute2019quantum,wu2021strong} and have shown promise in running proof-of-principle quantum machine learning applications~\cite{li2015experimental,bartkiewicz2020experimental,blank2020quantum,guan2020quantum,peters2021machine,wu2021application,havlivcek2019supervised,johri2020nearest,huang2020experimental,hubregtsen2021training,kusumoto2021experimental,dutta2021realization}. 
\revC{However, currently available quantum computers are at least 6 orders of magnitude orders slower than classical computers. Furthermore, running quantum computers is comparatively expensive, necessitating methods to reduce quantum resources above all else. Thus, it is important to develop better methods to run and benchmark noisy quantum computers. 
Here, several bottlenecks limit quantum hardware for machine learning in practice.
First, the quantum cost of measuring quantum kernels with conventional methods scales quadratically with the size of the training dataset~\cite{lloyd2020quantum}.} This quadratic scaling is a severe restriction, as commonly machine learning relies on large amounts of data.
Second, the data has to be encoded into the quantum computer in an efficient manner and generate a useful quantum kernel. Various encodings have been proposed~\cite{perez2020data,schuld2021quantum}, however the number of features is often limited by the number of qubits~\cite{havlivcek2019supervised,wu2021application} or the quantum kernel is characterized only in a heuristic manner. 
Finally, the inherent noise of quantum computers limits the quality of the experimental results. Error mitigation has been proposed to reduce the effect noise~\cite{temme2017error}, however in general this requires a large amount of additional quantum computing resources~\cite{endo2021hybrid}.

Here, we use randomized measurements to calculate quantum kernels. \revC{The quantum computing time scales linearly and the classical post-processing time quadratically with the size of the dataset.  While our method scales in general exponentially in the number of qubits, compared to other methods a substantially lower number of measurements is needed for intermediate-sized quantum computers of about ten qubits.}
Additionally, we can reuse the collected measurement data to effectively mitigate the noise of quantum computers.
To efficiently load high-dimensional data into the quantum computer, we apply an encoding that scales linearly with the depth of parameterized quantum circuits (PQCs). The resulting quantum kernel is characterized with the quantum Fisher information metric (QFIM) and can be approximately described by the radial basis function kernel. We introduce the natural PQC (NPQC) with an exactly known QFIM and demonstrate its usefulness for quantum machine learning.
We implement our approach on the IBM quantum computer to classify handwritten images of digits with high accuracy. We experimentally demonstrate further speedups by parallelizing quantum computational tasks between different quantum computers. 
\revC{With our approach, currently available quantum computers can process larger datasets containing ten thousands of entries within a feasible time, extending the range of quantum machine learning algorithms that can be run in practice.}

\section{Support vector machine}
%\emph{Support vector machine---}
\begin{figure*}[htbp]
	\centering
	\subfigimg[width=0.95\textwidth]{}{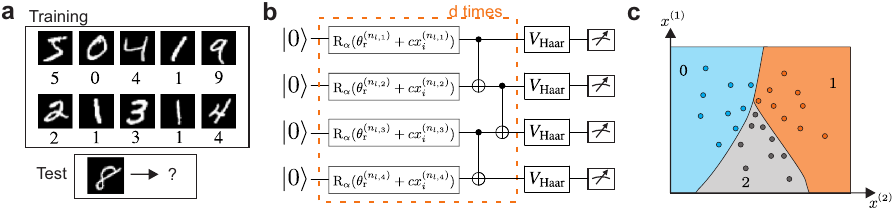}
	\caption{
	\idg{a} Supervised learning to classify images of handwritten digits. By learning from a training set of labeled images, our goal is to identify previously unseen test data correctly. The support vector machine (SVM) learns using a kernel (\eqref{eq:kernel}) which is a measure of distance between the data.
	\idg{b} \revB{We learn a dataset of $L$ images with $i=\{1,\dots,L\}$, each with $M$ pixels $n=\{1,\dots, M\}$, where we denote each pixel with $\boldsymbol{x}_i^{(n)}\in\mathbb{R}$. For the $i$th image, we encode the $M$-dimensional feature vector $\boldsymbol{x}_i$ into a parameterized quantum circuit (PQC) with an $M$-dimensional parameter vector $\boldsymbol{\theta}_i$. The PQC has $N$ qubits and $d$ layers of parameterized single qubit rotations and two-qubit entangling gates. We encode the $n_{l,k}\equiv n$ entry of the feature vector into a single qubit rotation acting on qubit $k$ and layer $l$ via $\theta_i^{(n)}=\theta_\text{r}^{(n)}+cx_i^{(n)}$ (\eqref{eq:encoding}), where $\boldsymbol{\theta}_\text{r}$ is a fixed reference parameter and $c$ a scaling factor.} The number of encoded features scales linearly with $N$ and $d$.
	The kernel (\eqref{eq:kernel}) is characterized by the quantum Fisher information metric (QFIM) $\mathcal{F}(\boldsymbol{\theta}_\text{r})$ and can be approximately described by the radial basis function kernel (\eqref{eq:rbf_kernel}).
	We calculate the quantum kernel by measuring the PQC in randomized local bases of Haar random unitaries $V_\text{Haar}$. 
	\idg{c} The SVM trained with the quantum kernel draws the decision boundaries (here shown for a two-dimensional feature vector space and three possible digits) that classify each feature vector to its corresponding label. 
	}
	\label{fig:sketch}
\end{figure*}
Our goal is to classify unlabeled test data by learning from labeled training data as shown in Fig.\ref{fig:sketch}a. 
\revC{The dataset for the supervised learning task $\{\{\boldsymbol{x}_i,y_i\}\}_{i=1}^L$ contains in total $L$ items.} The $i$-th data item is described by a $M$-dimensional feature vector $\boldsymbol{x}_i$ and corresponding label $y_i$. Label $y_i$ belongs to $C$ possible classes, while the feature vector $\boldsymbol{x}_i=\{\boldsymbol{x}_i^{(n)}\}_{n=1}^M$ consists of $M$ real-valued entries.
To learn and classify data, we use a kernel $K(\boldsymbol{x}_i,\boldsymbol{x}_j)$ that is a measure of distance between feature vectors $\boldsymbol{x}_i$ and $\boldsymbol{x}_j$~\cite{schuld2018supervised}. The kernel corresponds to an embedding of the $M$-dimensional data into a higher-dimensional space, where analysis of the data becomes easier~\cite{scholkopf2018learning}. In quantum kernel learning, we embed the data into the high-dimensional Hilbert space of the quantum computer and use it to calculate the kernel (see  Fig.\ref{fig:sketch}b).
With the kernels, we train a support vector machine (SVM) to find hyperplanes that separate two classes of data (see Fig.\ref{fig:sketch}c). 
The SVM is optimized using the kernels of the training dataset with a semidefinite program that can be efficiently solved with classical~\cite{wolkowicz2012handbook} or quantum computers~\cite{brandao2017quantum,bharti2021nisq}.
\begin{equation}\label{eq:SVM}
\text{max}_{\boldsymbol{\alpha}} \sum_i\alpha_i-\frac{1}{2}\sum_{i,j}y_iy_j\alpha_i\alpha_jK(\boldsymbol{x}_i,\boldsymbol{x}_j)
\end{equation}
subject to the conditions $\sum_i\alpha_iy_i=0$ and $\alpha_i\ge0$. After finding the optimal weights $\boldsymbol{\alpha}^*$, the SVM predicts the class of a feature vector $\boldsymbol{\eta}$ as
$y_i^\text{pred}=\text{sign}(\sum_i\alpha_i^*y_iK(\boldsymbol{x}_i,\boldsymbol{\eta})+b)$, where $b$ is calculated from the weights. One can extend this approach to distinguish $C$ classes by solving $C$ SVMs that separate each class from all other classes.

The power of the SVM highly depends on a good choice of kernel $K(\boldsymbol{x}_i,\boldsymbol{x}_j)$, such that it captures the essential features of the dataset. 
In the following, we propose a powerful class of quantum kernels that can be implemented with currently available quantum computers. Then, we show how to compute kernels for large datasets and mitigate the noise inherent in real quantum devices.

\section{Encoding}
%\emph{Encoding---}
A crucial question is how to efficiently encode a high-dimensional feature vector into a quantum computer while providing a useful kernel for machine learning. 
We encode the $M$-dimensional feature vector $\boldsymbol{x}_i$ as $M$-dimensional parameter $\boldsymbol{\theta}_i$ of a PQC via
\begin{equation}\label{eq:encoding}
\boldsymbol{\theta}_i=\boldsymbol{\theta}_\text{r}+c\boldsymbol{x}_i\,,
\end{equation}
where $c$ is a scaling constant and $\boldsymbol{\theta}_\text{r}$ the reference parameter. 
As shown in Fig.\ref{fig:sketch}b, we use hardware efficient PQCs with $N$ qubits and $d$ layers of unitaries for the encoding~\cite{kandala2017hardware}. \revB{The $l$th layer is composed of a product of parameterized single qubit rotations $R_{l,k}(\boldsymbol{\theta}_i^{(n_{l,k})})$ acting on qubit $k$ and non-parameterized entangling gates $W_l$ that generate the quantum state $\ket{\psi(\boldsymbol{\theta}_i)} =\prod_{l=1}^d W_l(\prod_{k} R_{l,k}(\boldsymbol{\theta}_i^{(n_{l,k})}))\ket{0}^{\otimes N}$. }

Our choice of quantum kernel measures the distance between two encoding states as given by the fidelity between $\rho(\boldsymbol{\theta}_i)$ and $\rho(\boldsymbol{\theta}_j)$~\cite{schuld2021quantum,huang2021power}
\begin{equation}\label{eq:kernel}
K(\boldsymbol{\theta}_i,\boldsymbol{\theta}_j)=\text{Tr}(\rho(\boldsymbol{\theta}_i)\rho(\boldsymbol{\theta}_j))\,,
\end{equation}
which for pure states $\rho(\boldsymbol{\theta}_i)=\ket{\psi(\boldsymbol{\theta}_i)}\bra{\psi(\boldsymbol{\theta}_i)}$ reduces to $K(\boldsymbol{\theta}_i,\boldsymbol{\theta}_j)=\abs{\braket{\psi(\boldsymbol{\theta}_i)}{\psi(\boldsymbol{\theta}_j)}}^2$.

We can formalize the expressive power of our encoding with the QFIM $\mathcal{F}(\boldsymbol{\theta})$, which is a $M\times M$ dimensional positive-semidefinite matrix that provides information about the kernel in the proximity of $\boldsymbol{\theta}$~\cite{haug2021capacity}.
For a pure state $\ket{\psi}=\ket{\psi(\boldsymbol{\theta})}$ it is given by
$\mathcal{F}_ {ij}(\boldsymbol{\theta})=4[\braket{\partial_i \psi}{\partial_j \psi}-\braket{\partial_i \psi}{\psi}\braket{\psi}{\partial_j \psi}]$,
where $\partial_j\ket{\psi}$ is the gradient in respect to the $j$-th element of $\boldsymbol{\theta}$~\cite{meyer2021fisher}.
In the limit $c\rightarrow0$ of encoding \eqref{eq:encoding}, the kernel of a pure quantum state can be written as
\begin{equation}\label{eq:fidelity_QFIM}
K(\boldsymbol{\theta}_\text{r},\boldsymbol{\theta}_\text{r}+c\boldsymbol{x}_i)= 1-\frac{c^2}{4}\boldsymbol{x}_i^{\text{T}}\mathcal{F}(\boldsymbol{\theta}_\text{r})\boldsymbol{x}_i=1-\frac{c^2}{4}\sum_{k=1}^M\lambda_k g_k\,,
\end{equation}
where $\lambda_k$ is the $k$-th eigenvalue of the QFIM $\mathcal{F}(\boldsymbol{\theta}_\text{r})$ and $g_k=\vert\langle \boldsymbol{x}_i,\boldsymbol{\mu}_k\rangle\vert^2$ is the inner product of the feature vector $\boldsymbol{x}_i$ and the $k$-th eigenvector $\boldsymbol{\mu}_k$ of $\mathcal{F}(\boldsymbol{\theta}_\text{r})$. 
$R=\text{rank}(\mathcal{F})$ (the number of non-zero eigenvalues) of $\mathcal{F}(\boldsymbol{\theta}_\text{r})$ is an important measure of the properties of the PQC and the encoding~\cite{haug2021capacity}.
The $M-R$ eigenvectors $\boldsymbol{\mu}_k$ with $\lambda_k=0$ have no effect on the kernel with $K(\boldsymbol{\theta},\boldsymbol{\theta}+c\boldsymbol{\mu}_k)=1$. Thus, feature vectors $\boldsymbol{x}_q\in\text{span}\{\boldsymbol{\mu}_1,\dots,\boldsymbol{\mu}_{M-R}\}$ that lie in the space of eigenvectors with eigenvalue zero cannot be distinguished using the kernel as they have the same value $K(\boldsymbol{\theta},\boldsymbol{\theta}+c\boldsymbol{x}_q)=1$. 
Further, the size of the eigenvalues $\lambda_k$ determines how strongly the kernel changes in direction $\boldsymbol{\mu}_k$ of the feature space. 
By appropriately designing the QFIM as the weight matrix of the kernel, generalizing from data could be greatly enhanced~\cite{haug2021capacity,schuld2021quantum,banchi2021generalization}. For example, the feature subspace with eigenvalue 0 could be engineered such that it coincides with data that belongs to a particular class. Conversely, features that strongly differ between different classes could be tailored to have large eigenvalues such that they can be easily distinguished~\cite{banchi2021generalization}.
For a PQC with $N$ qubits the rank is upper bounded by $R\le 2^{N+1}-2$, which is the maximal number of features that can be reliably distinguished by the kernel~\cite{haug2021capacity}.

\begin{figure*}[t]
	\centering	
	\subfigimg[width=0.35\textwidth]{a}{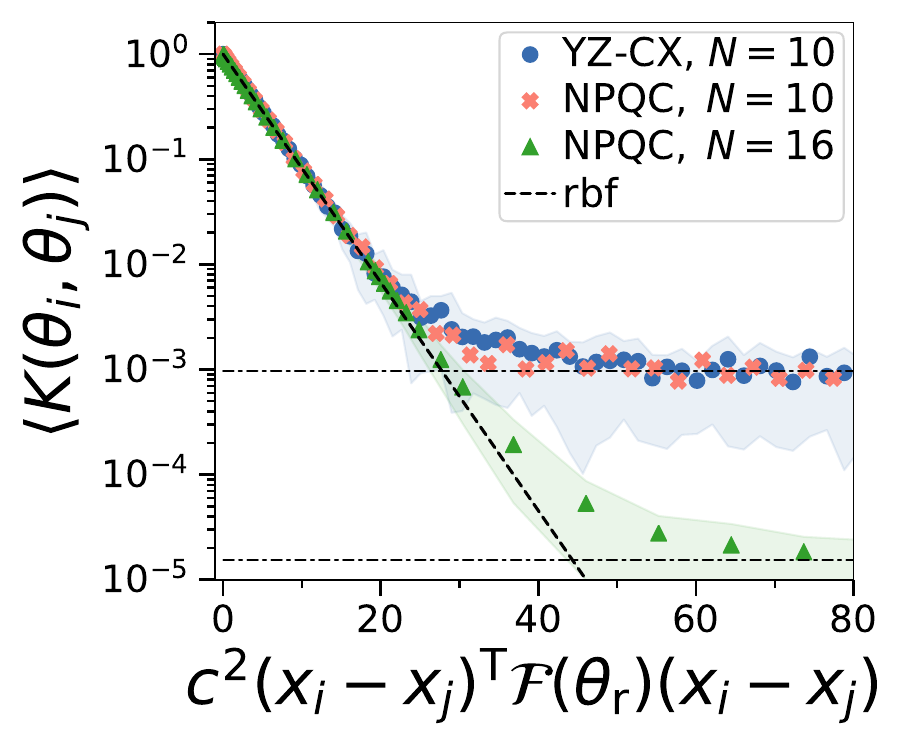} \hspace{12pt}
	\subfigimg[width=0.35\textwidth]{b}{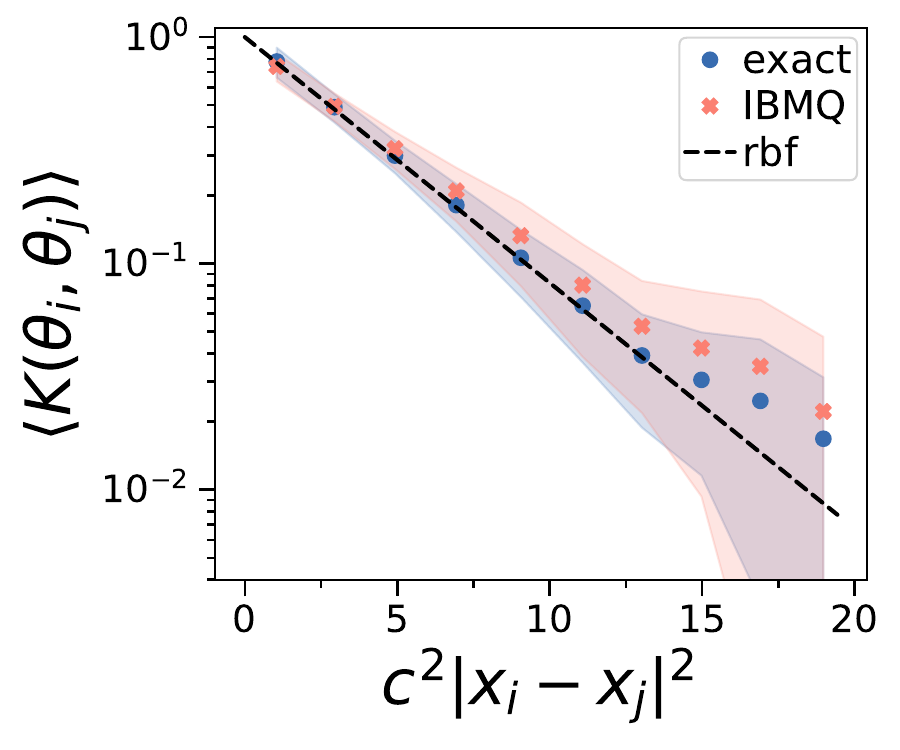}
	\caption{\idg{a} Simulated kernel $K(\boldsymbol{\theta}_i,\boldsymbol{\theta}_j)$ as function of 	$(\boldsymbol{x}_i-\boldsymbol{x}_j)^\text{T}\mathcal{F}(\boldsymbol{\theta}_\text{r})(\boldsymbol{x}_i-\boldsymbol{x}_j)$, which is the distance in feature space weighed with the QFIM $\mathcal{F}(\boldsymbol{\theta}_\text{r})$.
    \revB{The feature vectors $\boldsymbol{x}_i$ are sampled uniformly from $[-1,1]^{M}$, then the norm of $\boldsymbol{x}_i$ is rescaled to a randomly chosen $\vert\boldsymbol{x}_i\vert$ and encoded with \eqref{eq:encoding} into the PQC.}
	We show two types of hardware efficient PQCs, namely YZ-CX PQC and NPQC (see Appendix~\ref{sec:PQC}). Shaded area is the standard deviation of the kernel. The quantum kernels are well approximated by radial basis function kernels (rbf, dashed line, \eqref{eq:rbf_kernel}) until reaching very small values $K_\text{min}={2^{-N}}$ (dash-dotted lines). PQCs have $N=10$ qubits, $d=10$ layers and we average over 50 random feature vectors. 
	\idg{b} Experimental kernel $K(\boldsymbol{\theta}_i,\boldsymbol{\theta}_j)$ as function of distance between the feature vectors. We encode uniformly sampled $\boldsymbol{x}_i$ in the NPQC with the QFIM $\mathcal{F}(\boldsymbol{\theta}_\text{r})=I$. The quantum kernel generated by theory (blue dots) and via experimental results with IBM quantum computer (orange crosses) follows approximately the isotropic radial basis function kernel ($K(\boldsymbol{\theta}_i,\boldsymbol{\theta}_j)=\exp(-\frac{c^2}{4}\vert \boldsymbol{x}_i-\boldsymbol{x}_j\vert^2)$, black line). Shaded area is standard deviation of the kernel. The NPQC has $N=8$ qubits, $M=36$ features and $d=4$ layers. Experimental results from \emph{ibmq\_guadalupe} were performed with $r=50$ randomly chosen measurement settings, $s=8192$ measurement samples and error mitigation with \eqref{eq:kernel_mitigate}.
	}
	\label{fig:kernel}
\end{figure*}

It has been recently shown that the kernel of pure quantum states of hardware efficient PQCs can be approximated as Gaussian or radial-basis function kernels~\cite{haug2021optimal}, which are one of the most popular non-linear kernels with wide application in various machine learning methods~\cite{goodfellow2016deep}. Specifically, for small enough $c$ with the encoding~\eqref{eq:encoding}, we can approximately describe the quantum kernel as
\begin{equation}\label{eq:rbf_kernel}
K(\boldsymbol{\theta}_i,\boldsymbol{\theta}_j)\approx\text{exp}[-\frac{c^2}{4}(\boldsymbol{x}_i-\boldsymbol{x}_j)^{\text{T}}\mathcal{F}(\boldsymbol{\theta}_\text{r})(\boldsymbol{x}_i-\boldsymbol{x}_j)]\,,
\end{equation}
which is the radial basis function kernel with the QFIM as weight matrix $\mathcal{F}(\boldsymbol{\theta}_\text{r})$~\cite{haug2021optimal}. 
While for general PQCs the QFIM is a priori not known, a type of PQC called NPQC has the special property that the QFIM takes a simple form with $\mathcal{F}(\boldsymbol{\theta}_\text{r})=I$, where $I$ is the identity matrix and $\boldsymbol{\theta}_\text{r}$ a particular reference parameter, which we will choose in the following for the NPQC (see~\cite{haug2021natural} and Appendix~\ref{sec:PQC}). The NPQC forms an approximate isotropic radial basis function kernel that can serve as a well characterised basis for quantum machine learning. 
We also study another commonly used type of hardware efficient circuit (YZ-CX PQC) composed of single qubit rotations and CNOT gates arranged in a one-dimensional nearest-neighbor chain with a non-trivial QFIM $\mathcal{F}(\boldsymbol{\theta}_\text{r})\ne I$. \revB{ For the YZ-CX PQC we choose a randomly drawn $\boldsymbol{\theta}_\text{r}$, we find that the overall performance is nearly independent of the choice.}

Further details on the NPQC and YZ-CX PQC are shown in the Appendix~\ref{sec:PQC}.
\revA{The scaling factor $c$ controls the scale of the resulting values of the quantum kernel. Too small kernel values can impede learning as the model becomes too constrained. We can restrict the kernel from below $K_\text{min}<K(\boldsymbol{\theta}_i,\boldsymbol{\theta}_j)$ for all $i,j$  by choosing $c$ as
\begin{equation}
c^2<\frac{-4\log(K_\text{min})}{\text{min}_{i,j}(\boldsymbol{x}_i-\boldsymbol{x}_j)^{\text{T}}\mathcal{F}(\boldsymbol{\theta}_\text{r})(\boldsymbol{x}_i-\boldsymbol{x}_j)}\,.
\end{equation}}

\section{Measurement}
%\emph{Measurement---}
We calculate the $L$ quantum kernels using randomized measurements~\cite{elben2019statistical,elben2020cross,zhu2021crossplatform} by measuring quantum states in $r$ randomly chosen single qubit bases. We first choose $r$ sets $n=\{1,\dots,r\}$ of transformations $V^{(n)}=\otimes_{k=1}^N V_{k}^{(n)}$, composed of random single qubit rotations $V_{k}^{(n)}$ drawn according to the Haar measure $\mathbb{SU}(2)$ acting on each qubit $k$. Then, we prepare the quantum state $\rho(\boldsymbol{\theta}_i)$ and rotate into a random basis $V^{(n)}\rho(\boldsymbol{\theta}_i) {V^{(n)}}^\dagger$. Then, we measure $s$ samples of the rotated state in the computational basis and estimate the probability $P_i^{(n)}(v_q)$ of measuring the computational basis state $v\in\{0,1\}^{N}$ for state $\rho(\boldsymbol{\theta}_i)$ and transformation $n$. This procedure is repeated for the $r$ transformations and $L$ quantum states.
The kernel $K_\text{b}(\boldsymbol{\theta}_i,\boldsymbol{\theta}_j)=\text{Tr}(\rho(\boldsymbol{\theta}_i)\rho(\boldsymbol{\theta}_j))$ via randomized measurements is then calculated as~\cite{elben2020cross}
\begin{align*}
    K_\text{b}(\boldsymbol{\theta}_i,\boldsymbol{\theta}_j)&=\text{Tr}(\rho(\boldsymbol{\theta}_i)\rho(\boldsymbol{\theta}_j))=\\
    &\sum_{v,v'} (-2)^{-D(v,v')}\sum_{n=1}^rP_i^{(n)}(s)P_j^{(n)}(s')\,,\numberthis\label{eq:kernel_sample}
\end{align*}
where $D(v,v')$ is the Hamming distance that counts the number of bits that differ between the computational states $v$ and $v'$.

\revB{To measure all entries of the kernel, we perform $N_\text{R}=srL$ measurements in total. The error $\Delta K$ of estimating a single kernel entry scales as $\Delta K \propto 1/(s\sqrt{r})$~\cite{elben2020cross}. Thus, for a fixed error it is beneficial to choose the number of bases $r$ to a relatively small number compared to $s$. Note that for sufficient accuracy a minimal number of $r$ is needed which increases with $N$.  Overall, the number of measurements needed to estimate the kernel scales as $N_\text{R}\propto 2^{aN}L$, with a factor $a\lesssim 1$ that depends on the type of state being measured~\cite{elben2019statistical,elben2020cross} and can be improved by importance sampling~\cite{rath2021importance}. While for large $N$, the exponential measurement cost is prohibitive, for intermediate qubit number on the order of ten qubits the measurement cost is moderate.
With our method, the number of measurements needed to determine the full kernel matrix scales only linearly with the dataset size $N_\text{R}\propto L$, a quadratic speedup in contrast to other methods.}
Other commonly used measurement strategies such as the swap test~\cite{buhrman2001quantum,nguyen2021experimental} or the inversion test~\cite{havlivcek2019supervised,peters2021machine} have to explicitly prepare both states $\rho(\boldsymbol{\theta}_i)$ and $\rho(\boldsymbol{\theta}_j)$ on the quantum computer. Thus, they scale unfavorably with the square $N_\text{R}\propto L^2$ of the dataset size (see Appendix~\ref{sec:meas_kernel}). \revB{While randomized measurements requires an overhead compared to standard methods, we find that for relatively small datasets, $L>21$, randomized measurement requires less measurements for our experimental parameters (see Appendix~\ref{sec:cost}). For $L=10^3$, we find that randomized measurement requires a factor 100 lower number of measurements compared to the parameters used in previous works. 
A further advantage is found in error mitigation. For standard measurement methods on noisy quantum computers, error mitigation adds a substantial cost to the measurement budget~\cite{endo2021hybrid}. In contrast, randomized measurement can mitigate errors without further measurement cost as we show in the following.}

\section{Error mitigation}
%\emph{Error mitigation---}
In general, quantum computers are affected by noise, which will turn the prepared pure quantum state into a mixed state and may negatively affect the capability to learn.
For depolarizing noise, we can use the information gathered in the process to mitigate its effect and infer the noiseless value of the kernel.

For global depolarizing noise, with a probability $p_i$ the pure quantum state $\ket{\psi(\boldsymbol{\theta}_i)}$ is replaced with the completely mixed state $\rho_\text{m}=I/2^N$, where $I$ is the identity matrix. The resulting quantum state is the density matrix $\rho(\boldsymbol{\theta}_i)=(1-p_i)\ket{\psi(\boldsymbol{\theta}_i)}\bra{\psi(\boldsymbol{\theta}_i)}+p_i(2-p_i)\rho_\text{m}$.
The purity can be determined from the randomized measurements
$\text{Tr}(\rho(\boldsymbol{\theta}_i)^2)=K_\text{b}(\boldsymbol{\theta}_i,\boldsymbol{\theta}_i)=(1-p_i)^2+\frac{p_i}{2^N}$ by reusing the same data used to compute the kernel entries.
Using these purities, the depolarization probability $p_i$ can be calculated by solving a quadratic equation~\cite{vovrosh2021efficient,hubregtsen2021training}.
With $p_i$ and the measured kernel $K_\text{b}(\boldsymbol{\theta}_i,\boldsymbol{\theta}_j)$ affected by depolarizing noise, the mitigated kernel is approximated by
\begin{equation}\label{eq:kernel_mitigate}
    K_\text{m}(\boldsymbol{\theta}_i,\boldsymbol{\theta}_j)\approx \frac{K_\text{b}(\boldsymbol{\theta}_i,\boldsymbol{\theta}_j)}{\sqrt{\text{Tr}(\rho(\boldsymbol{\theta}_i)^2)\text{Tr}(\rho(\boldsymbol{\theta}_j)^2)}} \,.
\end{equation}
%\begin{equation}\label{eq:kernel_mitigate}
    %K_\text{m}(\boldsymbol{\theta}_i,\boldsymbol{\theta}_j)=\frac{K_\text{b}(\boldsymbol{\theta}_i,\boldsymbol{\theta}_j)-2^{-N}}{(1-p_i)(1-p_j)}+\frac{1}{2^{N}} \,,
%\end{equation}
%which simplifies for small $p_i$, $p_j$ to $K_\text{m}(\boldsymbol{\theta}_i,\boldsymbol{\theta}_j)\approx K_\text{b}(\boldsymbol{\theta}_i,\boldsymbol{\theta}_j)/\sqrt{\text{Tr}(\rho(\boldsymbol{\theta}_i)^2)\text{Tr}(\rho(\boldsymbol{\theta}_j)^2)}$. 

\section{Results}
%\emph{Results---}
We now proceed to numerically and experimentally demonstrate our methods.
First, we investigate the kernel of our encoding. 
In Fig.\ref{fig:kernel}a we numerically simulate~\cite{johansson2012qutip,yao} two types of hardware efficient PQCs (YZ-CX PQC and NPQC) and show that the quantum kernel is well described by a radial basis function kernel (\eqref{eq:rbf_kernel}, dashed line). The kernel diverges from the radial basis function kernel for exponentially small values of the kernel and reaches a plateau at $K_\text{min}=\frac{1}{2^N}$, which is the fidelity of Haar random states~\cite{mcclean2018barren}.
In Fig.\ref{fig:kernel}b, we experimentally measure the kernel of the NPQC with an IBM quantum computer (\emph{ibmq\_guadalupe}~\cite{ibmq-devices})  using randomized measurements and error mitigation (\eqref{eq:kernel_mitigate}). We find that the mean value of the kernel matches well with the isotropic radial basis function kernel. See Appendix~\ref{sec:ibm_details} for details on the experiment and Appendix~\ref{sec:exp_kernel} for results regarding the YZ-CX PQC.

\begin{figure}[t]
	\centering	
	\subfigimg[width=0.32\textwidth]{a}{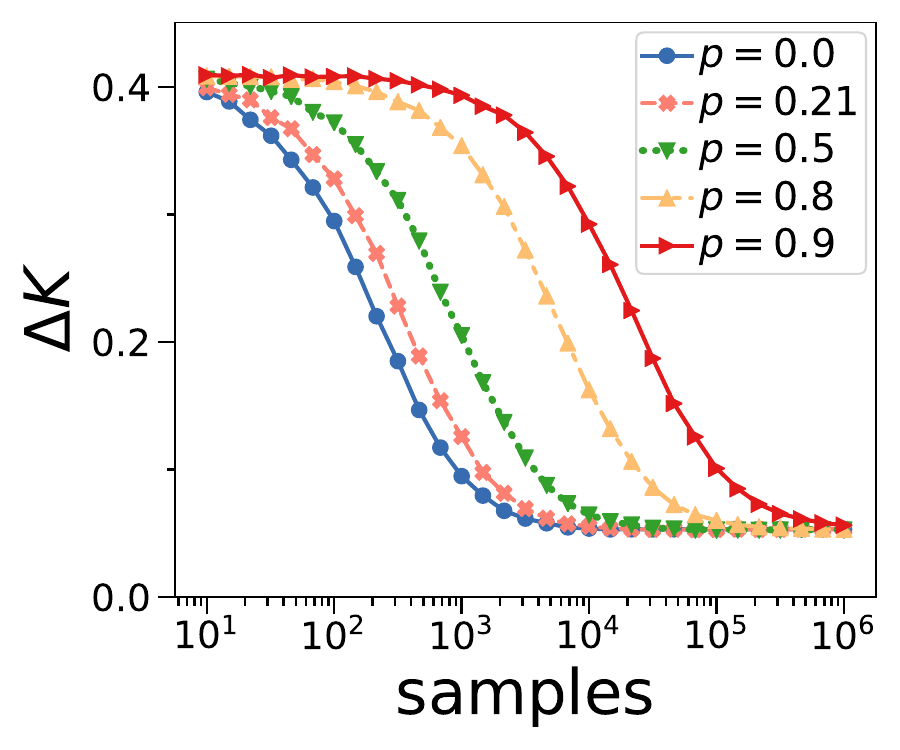} \\ % \hfill
	\subfigimg[width=0.32\textwidth]{b}{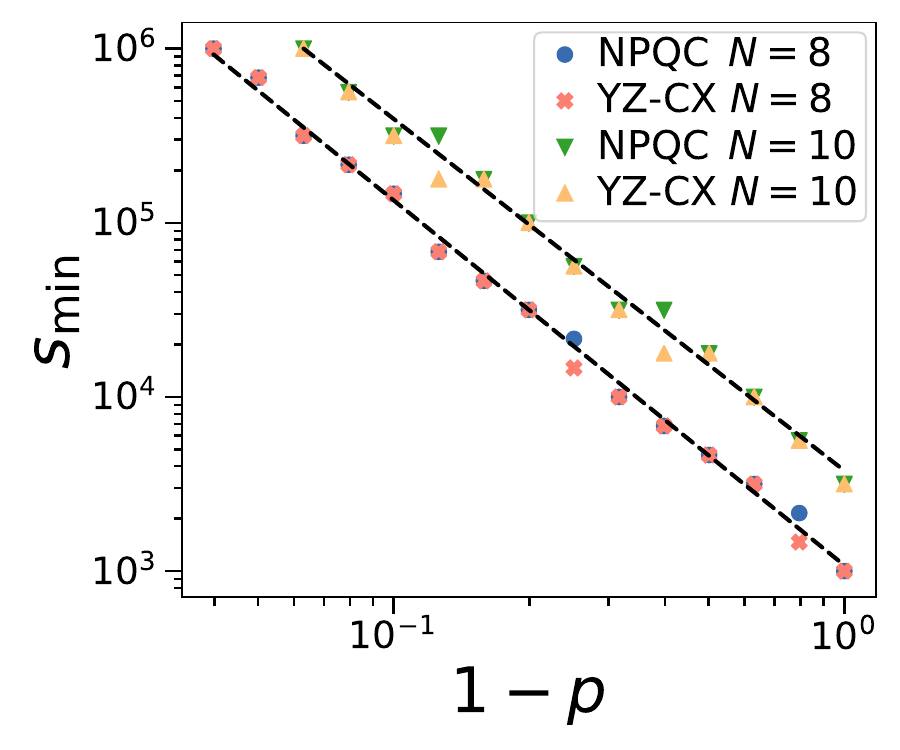}
	\caption{\idg{a} Average error for measuring the kernel with randomized measurements $\Delta K$ as function of number of measurement samples $s$ and the global depolarizing probability $p$. Simulation with $r=8$ measurement settings, $N=8$ qubits and the YZ-CX PQC.
	\idg{b} Minimal number of measurement samples $s_\text{min}$ needed to achieve an average error of at most $\Delta K<0.1$ for varying depolarizing noise $p$. Dashed line is the power law $s_\text{min}\propto (1-p)^{-2}$. Number of measurement settings is $r=8$ for $N=8$, and $r=16$ for $N=10$.
	}
	\label{fig:mitigation}
\end{figure}

Next we address the statistical error introduced by estimating the kernel using randomized measurements and global depolarizing noise $p$. In Fig.\ref{fig:mitigation}a we simulate the average error 
\begin{equation}\Delta K=\frac{2}{L(L-1)}\sum_{i=1}^L\sum_{j=i+1}^L\abs{K_\text{m}(\boldsymbol{\theta}_i,\boldsymbol{\theta}_j)-K(\boldsymbol{\theta}_i,\boldsymbol{\theta}_j)}
\end{equation}
of measuring the mitigated kernel $K_\text{m}(\boldsymbol{\theta}_i,\boldsymbol{\theta}_j)$ using randomized measurements with respect to its exact value $K(\boldsymbol{\theta}_i,\boldsymbol{\theta}_j)$ as function of number of measurement samples $s$. We find that there is a threshold of samples where the error becomes minimal. \revB{This threshold depends on the choice of the number of measurement settings $r$ and number of qubits $N$. We find that the choice $r=8$ provides sufficient accuracy for our experiments. }  
We are able to mitigate depolarizing noise to a noise-free level even for high $p$.
In Fig.\ref{fig:mitigation}b, we show the minimal number of samples $s_\text{min}$ required to measure the kernel with an average error of at most $\Delta K<0.1$ as function of depolarization noise $p$. The randomized measurement scheme works well even with substantial noise $p$, where we find a power law $s_\text{min}\propto (1-p)^{-2}$.

\begin{figure*}[t]
	\centering	
	\subfigimg[width=0.9\textwidth]{}{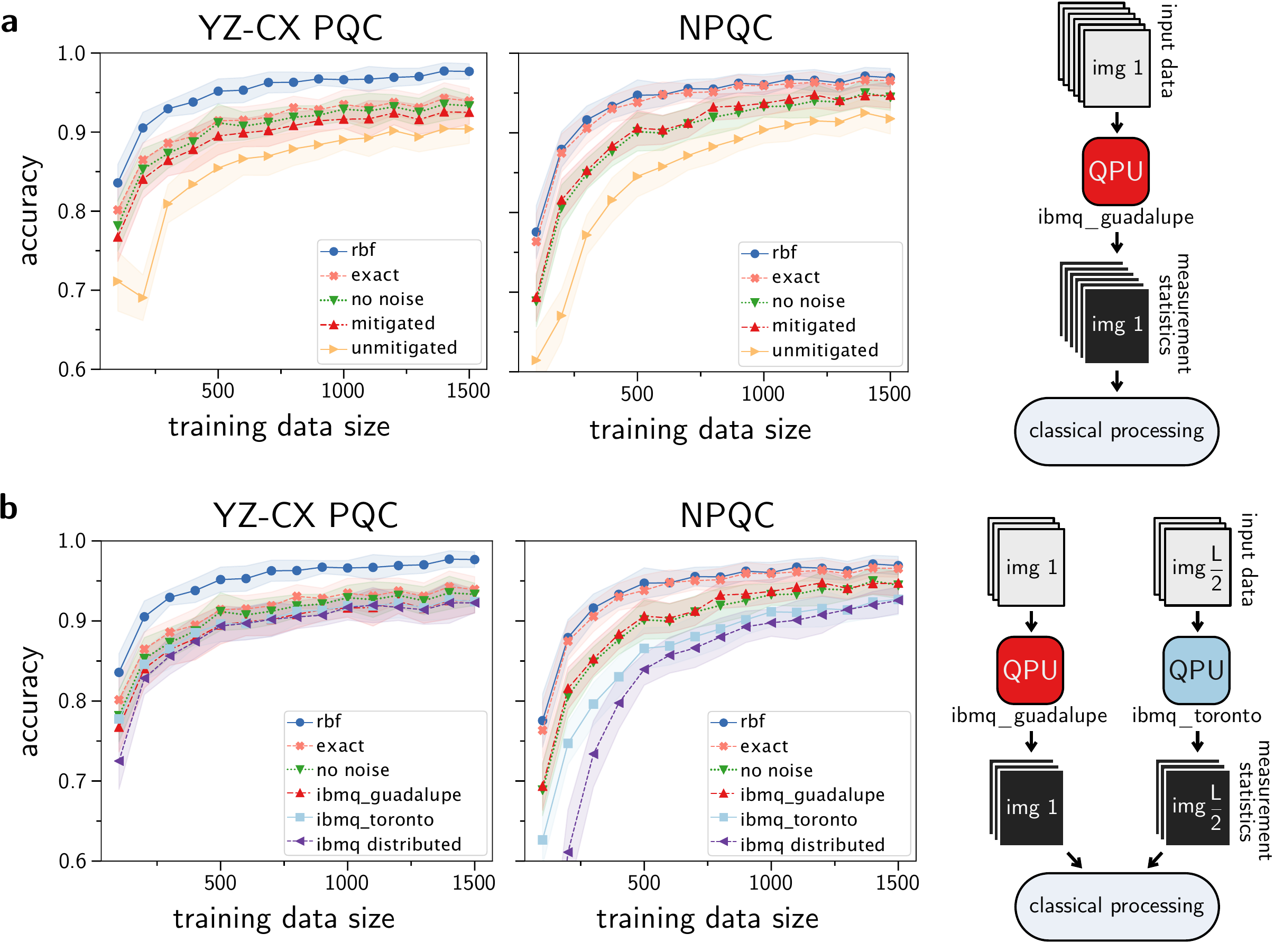} \hspace{12pt}
	\caption{Accuracy of classifying previously unseen handwritten digits correctly as function of the size of the training data. 
	\idg{a} SVM trained with experimental quantum kernel measured on a single quantum computer (\emph{ibmq\_guadalupe}) with randomized measurements using error mitigation (red, \eqref{eq:kernel_mitigate}) and no error mitigation (yellow). The shaded area is the standard deviation of the accuracy. \revA{As a classical baseline, we show the isotropic radial basis function kernel (blue).} Simulations of quantum kernels are the exact quantum kernel (orange) and noiseless simulation of randomized measurements (green).
	\idg{b} We distribute the measurements on two different quantum computers (\emph{ibmq\_guadalupe} and \emph{ibmq\_toronto}, purple curve) and post-process the combined measurement results with error mitigation. As reference, we show the accuracy of quantum kernel measured on a single quantum computer for \emph{ibmq\_guadalupe} (red) and \emph{ibmq\_toronto} (light blue).
    We encode the data into the YZ-CX PQC with $M=64$ features and the NPQC with $M=36$ features. 
    Experiments are performed using $s=8192$ measurement samples, $N=8$ qubits and $r=8$ randomized measurement settings. The test data contains $L_\text{test}=200$. \revB{To calculate mean and standard deviation of the accuracy, we randomly draw test and training data from the full dataset 20 times for each training data size.}
	}
	\label{fig:accuracy}
\end{figure*}

Now we assess the overall performance of our approach on a practical task. We learn to classify handwritten 2D images of digits ranging from 0 to 9. The dataset contains $L=1797$ images of $8 \times 8$ pixels, where each pixel has an integer value between 0 and 16~\cite{kaynak1995methods}. We map the image to $M=64$ dimensional feature vectors. For the YZ-CX PQC, we use all $M=64$ features, whereas for the NPQC we perform a principal component analysis to reduce it to $M=36$ features. 
We calculate the kernel of the full dataset and use a randomly drawn part of it as training data for optimizing the SVM with Scikit-learn~\cite{pedregosa2011scikit}. 
The accuracy of the SVM is defined as the percentage of correctly classified test data, which are $L_\text{test}=200$ images that have not been used for training. 
The dataset is rescaled using the training data such that each feature has mean value zero and its variance is given by $\frac{1}{\sqrt{M}}$. We encode the feature vectors $\boldsymbol{x}_i$ via \eqref{eq:encoding} with $c=1$, where for the YZ-CX PQC we choose $\boldsymbol{\theta}_\text{r}$ randomly and for the NPQC we define $\boldsymbol{\theta}_\text{r}$ such that the QFIM is given by $\mathcal{F}(\boldsymbol{\theta}_\text{r})=I$ (see Appendix~\ref{sec:PQC}).

In Fig.\ref{fig:accuracy}a, we classify the data by measuring the quantum kernel with a single quantum computer.
We plot the accuracy of classifying test data with the SVM against the size of the training data for the YZ-CX PQC and the NPQC. 
As a classical baseline, we show the radial basis function kernel (rbf). Further, we show a simulations of the exact quantum kernel (exact) and a noiseless simulation of the randomized measurements (noiseless). For experimental data, we use an IBM quantum computer (\emph{ibmq\_guadalupe}~\cite{ibmq-devices}, see Appendix~\ref{sec:ibm_details} for more details) to perform randomized measurements with error mitigation (mitigated) and without error mitigation (unmitigated).
The accuracy improves steadily with increased number of training data for all kernels. 
Our error mitigation scheme (\eqref{eq:kernel_mitigate}) substantially improves the accuracy of the SVM trained with experimental data to nearly the level of the noiseless simulation of the randomized measurements.
The randomized measurements have a lower accuracy compared to the exact quantum kernel as we use only a relatively small number $r$ of randomized measurement settings.
For the NPQC, the exact quantum kernel shows nearly the same accuracy as the classical radial basis function kernel, whereas for the YZ-CX PQC the quantum kernel performs slightly worse compared to the classical kernel, likely indicating that its QFIM does not optimally capture the structure of the data.
The depolarizing probability of the IBM quantum computer is estimated as $p\approx0.36$ for the NPQC and $p\approx0.39$ for the YZ-CX. \revB{To measure the kernel of the dataset with $L_\text{train}=1597$ and $L_\text{test}=200$, we require in total $N_\text{R}=s(L_\text{train}+L_\text{test})r\approx1.2\cdot 10^{8}$ measurements. For the inversion test, one would require $N_\text{R}=s_\text{c}L_\text{train}(L_\text{train}-1)/2+s_\text{c}L_\text{train}L_\text{test}\approx0.8\cdot 10^{10}$ experiments, where we have set the number of measurements per kernel entry to $s_\text{c}=5000$ as chosen in past experiments~\cite{peters2021machine}. Thus, we estimate that our method yields a reduction in total measurements by more than factor 60. We find that our method already yields a lower measurement cost for $L_\text{train}>21$ as shown in Appendix~\ref{sec:cost}.}

Finally, in Fig.\ref{fig:accuracy}b we distribute the measurements between two quantum computers. We split the dataset into two halves, where one half is measured using randomized measurements with \emph{ibmq\_guadalupe} and the other half with \emph{ibmq\_toronto}~\cite{ibmq-devices} (see Appendix~\ref{sec:ibm_details} for more details). The measurement outcomes from both machines are then combined for the post-processing on the classical computer to calculate the kernel matrix of the full dataset. Here, we also apply error mitigation. As reference, we also plot the accuracy achieved with a single quantum computer. For the YZ-CX PQC, we find nearly equal accuracy with the distributed and single quantum computer approach. For the NPQC, the accuracy of the distributed approach is slightly lower. 
The performance highly depends on the noise and calibration of the IBM quantum computers, which can fluctuate over time and highly depends when an experiment is performed. 
We attribute the lower performance of the distributed YZ-CX approach with a higher noise level present while the experiment was performed on \emph{ibmq\_toronto}.
As the randomized measurement method correlates measured samples, differences in the respective noise model of the two quantum computers can have a negative effect on the resulting quantum kernel.
In the Appendix~\ref{sec:training} and~\ref{sec:confusion}, we show the accuracy of the training data and the confusion matrices.

\section{Discussion}
%\emph{Discussion---}
\revB{Our work demonstrates a practical method to learn large datasets on noisy quantum computers with intermediate qubit numbers. Randomized measurement enables a linear scaling in dataset size $L$ and encodes high-dimensional data with number of features scaling linearly with quantum circuit depth $d$.}
We show our encoding can be characterized by the QFIM and its eigenvalues and eigenvectors~\cite{haug2021capacity}. As the behavior of the kernel is crucial for effectively learning and generalizing data, future work could design the QFIM to improve the capability of quantum machine learning models. We demonstrated the NPQC with a simple and exactly known QFIM, which could be a useful basis to study quantum machine learning on large quantum computers.

\revB{We encode the data in hardware efficient PQCs, which are known to be hard to simulate classically for large numbers of qubits~\cite{arute2019quantum}. This type of PQC has been used in quantum machine learning experiments~\cite{peters2021machine}. While sampling from these circuits is difficult to simulate on classical computers, we find that the quantum kernel closely follow the radial basis function kernel up to exponentially small kernel values~\cite{haug2021optimal}. Similarly, many other classes of quantum kernels have efficient classical representations~\cite{schreiber2022classical}. The resemblance with a classical kernel implies that these quantum kernels are unlikely to achieve an advantage over classical methods~\cite{huang2021power}.  However, we note that radial basis function type of kernels have been of interest in quantum optics~\cite{chatterjee2016generalized} and can serve as a reliable benchmark of quantum machine learning methods. Further, the non-trivial weight matrix $\mathcal{F}$ could be of independent interest in machine learning~\cite{musavi1992training}. }

%The relation of our quantum kernel and radial basis function kernels~\cite{haug2021optimal} gives us a strong indication that our encoding is at least as powerful as classical machine learning kernels and could be used to study the power of quantum machine learning~\cite{abbas2020power}. %However, we stress that the description as radial basis function kernels is only approximate and fails for small kernel values, which may hide possible quantum advantages~\cite{huang2021power}.

We mitigate the noise occurring in the quantum computer by using data sampled during the measurements of the kernel. 
We find that the number $s_\text{min}$ of measurement samples needed to mitigate depolarizing noise scales as $s_\text{min}\propto(1-p)^{-2}$, allowing us to extract kernels even from very noisy quantum computers. We successfully apply this model to mitigate the noise of the IBM quantum computer. While the noise model of quantum computers is known to be complicated involving multiple types of sources of noise, the depolarizing model we use is sufficient to mitigate the noise of kernels measured on IBM quantum computers~\cite{vovrosh2021efficient}. \revA{This may be the result of the randomized measurements leading to an insensitivity to fixed unitary noise channels.}
We note that noise induced errors can actually be beneficial to machine learning as the capability to generalize from data can improve with increasing noise~\cite{banchi2021generalization}.

\revB{In general, the number of measurements needed for the randomized measurement scheme scales exponentially with the number $N$ of qubits~\cite{elben2019statistical,elben2020cross}. This makes our method currently practical only for a lower number of qubits. However, various approaches could extend our method to larger qubit numbers. Importance sampling can reduce the number of measurements needed~\cite{rath2021importance}. For particular types of states an exponential reduction in cost has been observed. It would be worthwhile to study how importance sampling can improve the measurement cost for quantum machine learning.} In other settings adaptive measurements have been proposed to improve the scaling of measurement costs~\cite{garcia2021learning}, as well as other approaches such as shadow tomography~\cite{huang2020predicting}. The choice of an effective set of measurements could be included in the machine learning task as hyper-parameters to be optimised.
To reduce the number of qubits, one could combine our approach with quantum autoencoders to transform the encoding quantum state into a subspace with less qubits that captures the essential information of the kernel~\cite{romero2017quantum}. 
Alternatively, one could trace out most of the qubits of a many-qubit quantum state $\rho(\boldsymbol{\theta}_i)$ such that a subystem $A$ with a lower number of qubits remains. Then, randomized measurements can efficiently determine the kernel $\text{Tr}(\rho^A(\boldsymbol{\theta}_i)\rho^A(\boldsymbol{\theta}_j))$. 
It would be worthwhile to investigate the learning power of kernels generated from subsystems of quantum states that possess quantum advantage~\cite{liu2020rigorous,huang2021power}.

\revA{Randomized measurements process each of the $L$ quantum states of the dataset separately~\cite{elben2020cross}. The full kernel matrix $K(\boldsymbol{\theta}_i,\boldsymbol{\theta}_j)$ with $L^2$ elements is then constructed via classical post-processing using Eq.~\ref{eq:kernel_sample} where the randomized measurement data for state $| \psi ({\theta}_i) \rangle$, $| \psi ({\theta}_j) \rangle$ is reused to calculate each entry of the matrix. This gives us the resulting speedup in quantum computational time.
As a further advantage, our approach only requires preparing one quantum state at a time, reducing the number of gates by half compared to the inversion test or swap test.}
Further, we demonstrate how to achieve additional speedups by distributing measurements across different quantum computers.

\revC{The quantum computation time scales linearly with dataset size $L$ and provides a quadratic speedup compared to conventional measurement methods such as the inversion test or swap test. Note that the classical post-processing to construct the kernel still scales as $L^2$. However, we note that current quantum computers perform measurements at a rate of $\sim5\,\text{kHz}$~\cite{arute2019quantum,wu2021strong}, which is a factor $10^6$ slower than commonly available classical computers. Further, using quantum computers is very expensive compared to classical computation. Thus, the main bottleneck for quantum machine learning algorithms on current quantum hardware lies within the quantum part, while the classical part can be easily parallelized and distributed. Therefore, our work opens up benchmarking quantum machine learning with large datasets on intermediate-size quantum computers, which was impractical with previously known methods.
}

\revB{For our encoding~\eqref{eq:encoding}, at small distances the quantum kernel in parameter space can be described by the QFIM via~\eqref{eq:fidelity_QFIM}. We note this relation is general for any type of PQC.
The rank of the QFIM $\mathcal{F}$ indicates the number of independent directions in parameter space with \eqref{eq:fidelity_QFIM}. The maximal number of independent features $M_\text{max}$ that can be encoded via~\eqref{eq:encoding} is thus given by the rank of the QFIM, which is upper bounded by $\text{rank}(\mathcal{F})=M_\text{max}\le 2^{N+1}-2$~\cite{haug2021capacity}. 
Thus, even a modest number of qubits can represent a large number of parameters. The popular MNIST dataset~\cite{deng2012mnist} for classifying 2D images of handwritten digits has $28 \times 28$ pixels, which could be encoded in only $N=9$ qubits. }

Assuming $5\,\text{kHz}$ measurement rate, $s=8192$ measurement samples and $r=8$ measurement settings, our method can process the full MNIST training dataset with $L_\text{train}=60000$ entries in about 240 hours of quantum processing time of a single quantum computer. In contrast, the inversion or swap test would require at least 10 years with $s=1000$ samples on a quantum computer. 
\revB{With our scheme, we enable quantum machine learning with large datasets on intermediate-sized quantum computers. Future work could benchmark the performance of currently available quantum computers with datasets commonly used in classical machine learning.}

\section*{Data availability statement}
The code to reproduce the experimental results presented in this paper is available from~\cite{self2021qml} and %\href{https://github.com/chris-n-self/large-scale-qml}{https://github.com/chris-n-self/large-scale-qml}.
the experimental data is available from~\cite{self2021zenodoqml}.% \href{https://doi.org/10.5281/zenodo.5211695}{https://doi.org/10.5281/zenodo.5211695}.

\medskip
\begin{acknowledgments}
%{{\em Acknowledgements---}} %\noindent 
%\section{Acknowledgements}
We acknowledge discussions with Kiran Khosla and Alistair Smith. This work is supported by a Samsung GRC project and the UK Hub in Quantum Computing and Simulation, part of the UK National Quantum Technologies Programme with funding from UKRI EPSRC grant EP/T001062/1. We acknowledge the use of IBM Quantum services for this work. The views expressed are those of the authors, and do not reflect the official policy or position of IBM or the IBM Quantum team.
\end{acknowledgments}
\bibliography{QMLmeas}

\begin{thebibliography}{66}%
\makeatletter
\providecommand \@ifxundefined [1]{%
 \@ifx{#1\undefined}
}%
\providecommand \@ifnum [1]{%
 \ifnum #1\expandafter \@firstoftwo
 \else \expandafter \@secondoftwo
 \fi
}%
\providecommand \@ifx [1]{%
 \ifx #1\expandafter \@firstoftwo
 \else \expandafter \@secondoftwo
 \fi
}%
\providecommand \natexlab [1]{#1}%
\providecommand \enquote  [1]{``#1''}%
\providecommand \bibnamefont  [1]{#1}%
\providecommand \bibfnamefont [1]{#1}%
\providecommand \citenamefont [1]{#1}%
\providecommand \href@noop [0]{\@secondoftwo}%
\providecommand \href [0]{\begingroup \@sanitize@url \@href}%
\providecommand \@href[1]{\@@startlink{#1}\@@href}%
\providecommand \@@href[1]{\endgroup#1\@@endlink}%
\providecommand \@sanitize@url [0]{\catcode `\\12\catcode `\$12\catcode
  `\&12\catcode `\#12\catcode `\^12\catcode `\_12\catcode `\%12\relax}%
\providecommand \@@startlink[1]{}%
\providecommand \@@endlink[0]{}%
\providecommand \url  [0]{\begingroup\@sanitize@url \@url }%
\providecommand \@url [1]{\endgroup\@href {#1}{\urlprefix }}%
\providecommand \urlprefix  [0]{URL }%
\providecommand \Eprint [0]{\href }%
\providecommand \doibase [0]{http://dx.doi.org/}%
\providecommand \selectlanguage [0]{\@gobble}%
\providecommand \bibinfo  [0]{\@secondoftwo}%
\providecommand \bibfield  [0]{\@secondoftwo}%
\providecommand \translation [1]{[#1]}%
\providecommand \BibitemOpen [0]{}%
\providecommand \bibitemStop [0]{}%
\providecommand \bibitemNoStop [0]{.\EOS\space}%
\providecommand \EOS [0]{\spacefactor3000\relax}%
\providecommand \BibitemShut  [1]{\csname bibitem#1\endcsname}%
\let\auto@bib@innerbib\@empty
%</preamble>
\bibitem [{\citenamefont {Biamonte}\ \emph {et~al.}(2017)\citenamefont
  {Biamonte}, \citenamefont {Wittek}, \citenamefont {Pancotti}, \citenamefont
  {Rebentrost}, \citenamefont {Wiebe},\ and\ \citenamefont
  {Lloyd}}]{biamonte2017quantum}%
  \BibitemOpen
  \bibfield  {author} {\bibinfo {author} {\bibfnamefont {Jacob}\ \bibnamefont
  {Biamonte}}, \bibinfo {author} {\bibfnamefont {Peter}\ \bibnamefont
  {Wittek}}, \bibinfo {author} {\bibfnamefont {Nicola}\ \bibnamefont
  {Pancotti}}, \bibinfo {author} {\bibfnamefont {Patrick}\ \bibnamefont
  {Rebentrost}}, \bibinfo {author} {\bibfnamefont {Nathan}\ \bibnamefont
  {Wiebe}}, \ and\ \bibinfo {author} {\bibfnamefont {Seth}\ \bibnamefont
  {Lloyd}},\ }\bibfield  {title} {\enquote {\bibinfo {title} {Quantum machine
  learning},}\ }\href@noop {} {\bibfield  {journal} {\bibinfo  {journal}
  {Nature}\ }\textbf {\bibinfo {volume} {549}},\ \bibinfo {pages} {195--202}
  (\bibinfo {year} {2017})}\BibitemShut {NoStop}%
\bibitem [{\citenamefont {Schuld}\ and\ \citenamefont
  {Petruccione}(2018)}]{schuld2018supervised}%
  \BibitemOpen
  \bibfield  {author} {\bibinfo {author} {\bibfnamefont {Maria}\ \bibnamefont
  {Schuld}}\ and\ \bibinfo {author} {\bibfnamefont {Francesco}\ \bibnamefont
  {Petruccione}},\ }\href@noop {} {\emph {\bibinfo {title} {Supervised learning
  with quantum computers}}},\ Vol.~\bibinfo {volume} {17}\ (\bibinfo
  {publisher} {Springer},\ \bibinfo {year} {2018})\BibitemShut {NoStop}%
\bibitem [{\citenamefont {Schuld}\ and\ \citenamefont
  {Killoran}(2019)}]{schuld2019quantum}%
  \BibitemOpen
  \bibfield  {author} {\bibinfo {author} {\bibfnamefont {Maria}\ \bibnamefont
  {Schuld}}\ and\ \bibinfo {author} {\bibfnamefont {Nathan}\ \bibnamefont
  {Killoran}},\ }\bibfield  {title} {\enquote {\bibinfo {title} {Quantum
  machine learning in feature hilbert spaces},}\ }\href@noop {} {\bibfield
  {journal} {\bibinfo  {journal} {Physical review letters}\ }\textbf {\bibinfo
  {volume} {122}},\ \bibinfo {pages} {040504} (\bibinfo {year}
  {2019})}\BibitemShut {NoStop}%
\bibitem [{\citenamefont {Schuld}\ \emph {et~al.}(2021)\citenamefont {Schuld},
  \citenamefont {Sweke},\ and\ \citenamefont {Meyer}}]{schuld2021effect}%
  \BibitemOpen
  \bibfield  {author} {\bibinfo {author} {\bibfnamefont {Maria}\ \bibnamefont
  {Schuld}}, \bibinfo {author} {\bibfnamefont {Ryan}\ \bibnamefont {Sweke}}, \
  and\ \bibinfo {author} {\bibfnamefont {Johannes~Jakob}\ \bibnamefont
  {Meyer}},\ }\bibfield  {title} {\enquote {\bibinfo {title} {Effect of data
  encoding on the expressive power of variational quantum-machine-learning
  models},}\ }\href@noop {} {\bibfield  {journal} {\bibinfo  {journal}
  {Physical Review A}\ }\textbf {\bibinfo {volume} {103}},\ \bibinfo {pages}
  {032430} (\bibinfo {year} {2021})}\BibitemShut {NoStop}%
\bibitem [{\citenamefont {Lloyd}\ \emph {et~al.}(2020)\citenamefont {Lloyd},
  \citenamefont {Schuld}, \citenamefont {Ijaz}, \citenamefont {Izaac},\ and\
  \citenamefont {Killoran}}]{lloyd2020quantum}%
  \BibitemOpen
  \bibfield  {author} {\bibinfo {author} {\bibfnamefont {Seth}\ \bibnamefont
  {Lloyd}}, \bibinfo {author} {\bibfnamefont {Maria}\ \bibnamefont {Schuld}},
  \bibinfo {author} {\bibfnamefont {Aroosa}\ \bibnamefont {Ijaz}}, \bibinfo
  {author} {\bibfnamefont {Josh}\ \bibnamefont {Izaac}}, \ and\ \bibinfo
  {author} {\bibfnamefont {Nathan}\ \bibnamefont {Killoran}},\ }\bibfield
  {title} {\enquote {\bibinfo {title} {Quantum embeddings for machine
  learning},}\ }\href@noop {} {\bibfield  {journal} {\bibinfo  {journal}
  {arXiv:2001.03622}\ } (\bibinfo {year} {2020})}\BibitemShut {NoStop}%
\bibitem [{\citenamefont {Li}\ and\ \citenamefont {Deng}(2022)}]{li2021recent}%
  \BibitemOpen
  \bibfield  {author} {\bibinfo {author} {\bibfnamefont {Weikang}\ \bibnamefont
  {Li}}\ and\ \bibinfo {author} {\bibfnamefont {Dong-Ling}\ \bibnamefont
  {Deng}},\ }\bibfield  {title} {\enquote {\bibinfo {title} {Recent advances
  for quantum classifiers},}\ }\href@noop {} {\bibfield  {journal} {\bibinfo
  {journal} {Science China Physics, Mechanics \& Astronomy}\ }\textbf {\bibinfo
  {volume} {65}},\ \bibinfo {pages} {1--23} (\bibinfo {year}
  {2022})}\BibitemShut {NoStop}%
\bibitem [{\citenamefont {Liu}\ \emph {et~al.}(2021)\citenamefont {Liu},
  \citenamefont {Arunachalam},\ and\ \citenamefont {Temme}}]{liu2020rigorous}%
  \BibitemOpen
  \bibfield  {author} {\bibinfo {author} {\bibfnamefont {Yunchao}\ \bibnamefont
  {Liu}}, \bibinfo {author} {\bibfnamefont {Srinivasan}\ \bibnamefont
  {Arunachalam}}, \ and\ \bibinfo {author} {\bibfnamefont {Kristan}\
  \bibnamefont {Temme}},\ }\bibfield  {title} {\enquote {\bibinfo {title} {A
  rigorous and robust quantum speed-up in supervised machine learning},}\
  }\href@noop {} {\bibfield  {journal} {\bibinfo  {journal} {Nature Physics}\
  ,\ \bibinfo {pages} {1--5}} (\bibinfo {year} {2021})}\BibitemShut {NoStop}%
\bibitem [{\citenamefont {Huang}\ \emph
  {et~al.}(2021{\natexlab{a}})\citenamefont {Huang}, \citenamefont {Broughton},
  \citenamefont {Mohseni}, \citenamefont {Babbush}, \citenamefont {Boixo},
  \citenamefont {Neven},\ and\ \citenamefont {McClean}}]{huang2021power}%
  \BibitemOpen
  \bibfield  {author} {\bibinfo {author} {\bibfnamefont {Hsin-Yuan}\
  \bibnamefont {Huang}}, \bibinfo {author} {\bibfnamefont {Michael}\
  \bibnamefont {Broughton}}, \bibinfo {author} {\bibfnamefont {Masoud}\
  \bibnamefont {Mohseni}}, \bibinfo {author} {\bibfnamefont {Ryan}\
  \bibnamefont {Babbush}}, \bibinfo {author} {\bibfnamefont {Sergio}\
  \bibnamefont {Boixo}}, \bibinfo {author} {\bibfnamefont {Hartmut}\
  \bibnamefont {Neven}}, \ and\ \bibinfo {author} {\bibfnamefont {Jarrod~R}\
  \bibnamefont {McClean}},\ }\bibfield  {title} {\enquote {\bibinfo {title}
  {Power of data in quantum machine learning},}\ }\href@noop {} {\bibfield
  {journal} {\bibinfo  {journal} {Nature communications}\ }\textbf {\bibinfo
  {volume} {12}},\ \bibinfo {pages} {1--9} (\bibinfo {year}
  {2021}{\natexlab{a}})}\BibitemShut {NoStop}%
\bibitem [{\citenamefont {Huang}\ \emph {et~al.}(2022)\citenamefont {Huang},
  \citenamefont {Kueng}, \citenamefont {Torlai}, \citenamefont {Albert},\ and\
  \citenamefont {Preskill}}]{huang2021provably}%
  \BibitemOpen
  \bibfield  {author} {\bibinfo {author} {\bibfnamefont {Hsin-Yuan}\
  \bibnamefont {Huang}}, \bibinfo {author} {\bibfnamefont {Richard}\
  \bibnamefont {Kueng}}, \bibinfo {author} {\bibfnamefont {Giacomo}\
  \bibnamefont {Torlai}}, \bibinfo {author} {\bibfnamefont {Victor~V}\
  \bibnamefont {Albert}}, \ and\ \bibinfo {author} {\bibfnamefont {John}\
  \bibnamefont {Preskill}},\ }\bibfield  {title} {\enquote {\bibinfo {title}
  {Provably efficient machine learning for quantum many-body problems},}\
  }\href@noop {} {\bibfield  {journal} {\bibinfo  {journal} {Science}\ }\textbf
  {\bibinfo {volume} {377}},\ \bibinfo {pages} {eabk3333} (\bibinfo {year}
  {2022})}\BibitemShut {NoStop}%
\bibitem [{\citenamefont {Preskill}(2018)}]{preskill2018quantum}%
  \BibitemOpen
  \bibfield  {author} {\bibinfo {author} {\bibfnamefont {John}\ \bibnamefont
  {Preskill}},\ }\bibfield  {title} {\enquote {\bibinfo {title} {Quantum
  computing in the nisq era and beyond},}\ }\href@noop {} {\bibfield  {journal}
  {\bibinfo  {journal} {Quantum}\ }\textbf {\bibinfo {volume} {2}},\ \bibinfo
  {pages} {79} (\bibinfo {year} {2018})}\BibitemShut {NoStop}%
\bibitem [{\citenamefont {Bharti}\ \emph
  {et~al.}(2022{\natexlab{a}})\citenamefont {Bharti}, \citenamefont
  {Cervera-Lierta}, \citenamefont {Kyaw}, \citenamefont {Haug}, \citenamefont
  {Alperin-Lea}, \citenamefont {Anand}, \citenamefont {Degroote}, \citenamefont
  {Heimonen}, \citenamefont {Kottmann}, \citenamefont {Menke}, \citenamefont
  {Mok}, \citenamefont {Sim}, \citenamefont {Kwek},\ and\ \citenamefont
  {Aspuru-Guzik}}]{bharti2021noisy}%
  \BibitemOpen
  \bibfield  {author} {\bibinfo {author} {\bibfnamefont {Kishor}\ \bibnamefont
  {Bharti}}, \bibinfo {author} {\bibfnamefont {Alba}\ \bibnamefont
  {Cervera-Lierta}}, \bibinfo {author} {\bibfnamefont {Thi~Ha}\ \bibnamefont
  {Kyaw}}, \bibinfo {author} {\bibfnamefont {Tobias}\ \bibnamefont {Haug}},
  \bibinfo {author} {\bibfnamefont {Sumner}\ \bibnamefont {Alperin-Lea}},
  \bibinfo {author} {\bibfnamefont {Abhinav}\ \bibnamefont {Anand}}, \bibinfo
  {author} {\bibfnamefont {Matthias}\ \bibnamefont {Degroote}}, \bibinfo
  {author} {\bibfnamefont {Hermanni}\ \bibnamefont {Heimonen}}, \bibinfo
  {author} {\bibfnamefont {Jakob~S.}\ \bibnamefont {Kottmann}}, \bibinfo
  {author} {\bibfnamefont {Tim}\ \bibnamefont {Menke}}, \bibinfo {author}
  {\bibfnamefont {Wai-Keong}\ \bibnamefont {Mok}}, \bibinfo {author}
  {\bibfnamefont {Sukin}\ \bibnamefont {Sim}}, \bibinfo {author} {\bibfnamefont
  {Leong-Chuan}\ \bibnamefont {Kwek}}, \ and\ \bibinfo {author} {\bibfnamefont
  {Al\'an}\ \bibnamefont {Aspuru-Guzik}},\ }\bibfield  {title} {\enquote
  {\bibinfo {title} {Noisy intermediate-scale quantum algorithms},}\ }\href
  {\doibase 10.1103/RevModPhys.94.015004} {\bibfield  {journal} {\bibinfo
  {journal} {Rev. Mod. Phys.}\ }\textbf {\bibinfo {volume} {94}},\ \bibinfo
  {pages} {015004} (\bibinfo {year} {2022}{\natexlab{a}})}\BibitemShut
  {NoStop}%
\bibitem [{\citenamefont {Arute}\ \emph {et~al.}(2019)\citenamefont {Arute},
  \citenamefont {Arya}, \citenamefont {Babbush}, \citenamefont {Bacon},
  \citenamefont {Bardin}, \citenamefont {Barends}, \citenamefont {Biswas},
  \citenamefont {Boixo}, \citenamefont {Brandao}, \citenamefont {Buell} \emph
  {et~al.}}]{arute2019quantum}%
  \BibitemOpen
  \bibfield  {author} {\bibinfo {author} {\bibfnamefont {Frank}\ \bibnamefont
  {Arute}}, \bibinfo {author} {\bibfnamefont {Kunal}\ \bibnamefont {Arya}},
  \bibinfo {author} {\bibfnamefont {Ryan}\ \bibnamefont {Babbush}}, \bibinfo
  {author} {\bibfnamefont {Dave}\ \bibnamefont {Bacon}}, \bibinfo {author}
  {\bibfnamefont {Joseph~C}\ \bibnamefont {Bardin}}, \bibinfo {author}
  {\bibfnamefont {Rami}\ \bibnamefont {Barends}}, \bibinfo {author}
  {\bibfnamefont {Rupak}\ \bibnamefont {Biswas}}, \bibinfo {author}
  {\bibfnamefont {Sergio}\ \bibnamefont {Boixo}}, \bibinfo {author}
  {\bibfnamefont {Fernando~GSL}\ \bibnamefont {Brandao}}, \bibinfo {author}
  {\bibfnamefont {David~A}\ \bibnamefont {Buell}},  \emph {et~al.},\ }\bibfield
   {title} {\enquote {\bibinfo {title} {Quantum supremacy using a programmable
  superconducting processor},}\ }\href@noop {} {\bibfield  {journal} {\bibinfo
  {journal} {Nature}\ }\textbf {\bibinfo {volume} {574}},\ \bibinfo {pages}
  {505--510} (\bibinfo {year} {2019})}\BibitemShut {NoStop}%
\bibitem [{\citenamefont {Wu}\ \emph {et~al.}(2021{\natexlab{a}})\citenamefont
  {Wu}, \citenamefont {Bao}, \citenamefont {Cao}, \citenamefont {Chen},
  \citenamefont {Chen}, \citenamefont {Chen}, \citenamefont {Chung},
  \citenamefont {Deng}, \citenamefont {Du}, \citenamefont {Fan} \emph
  {et~al.}}]{wu2021strong}%
  \BibitemOpen
  \bibfield  {author} {\bibinfo {author} {\bibfnamefont {Yulin}\ \bibnamefont
  {Wu}}, \bibinfo {author} {\bibfnamefont {Wan-Su}\ \bibnamefont {Bao}},
  \bibinfo {author} {\bibfnamefont {Sirui}\ \bibnamefont {Cao}}, \bibinfo
  {author} {\bibfnamefont {Fusheng}\ \bibnamefont {Chen}}, \bibinfo {author}
  {\bibfnamefont {Ming-Cheng}\ \bibnamefont {Chen}}, \bibinfo {author}
  {\bibfnamefont {Xiawei}\ \bibnamefont {Chen}}, \bibinfo {author}
  {\bibfnamefont {Tung-Hsun}\ \bibnamefont {Chung}}, \bibinfo {author}
  {\bibfnamefont {Hui}\ \bibnamefont {Deng}}, \bibinfo {author} {\bibfnamefont
  {Yajie}\ \bibnamefont {Du}}, \bibinfo {author} {\bibfnamefont {Daojin}\
  \bibnamefont {Fan}},  \emph {et~al.},\ }\bibfield  {title} {\enquote
  {\bibinfo {title} {Strong quantum computational advantage using a
  superconducting quantum processor},}\ }\href@noop {} {\bibfield  {journal}
  {\bibinfo  {journal} {Physical review letters}\ }\textbf {\bibinfo {volume}
  {127}},\ \bibinfo {pages} {180501} (\bibinfo {year}
  {2021}{\natexlab{a}})}\BibitemShut {NoStop}%
\bibitem [{\citenamefont {Li}\ \emph {et~al.}(2015)\citenamefont {Li},
  \citenamefont {Liu}, \citenamefont {Xu},\ and\ \citenamefont
  {Du}}]{li2015experimental}%
  \BibitemOpen
  \bibfield  {author} {\bibinfo {author} {\bibfnamefont {Zhaokai}\ \bibnamefont
  {Li}}, \bibinfo {author} {\bibfnamefont {Xiaomei}\ \bibnamefont {Liu}},
  \bibinfo {author} {\bibfnamefont {Nanyang}\ \bibnamefont {Xu}}, \ and\
  \bibinfo {author} {\bibfnamefont {Jiangfeng}\ \bibnamefont {Du}},\ }\bibfield
   {title} {\enquote {\bibinfo {title} {Experimental realization of a quantum
  support vector machine},}\ }\href@noop {} {\bibfield  {journal} {\bibinfo
  {journal} {Physical review letters}\ }\textbf {\bibinfo {volume} {114}},\
  \bibinfo {pages} {140504} (\bibinfo {year} {2015})}\BibitemShut {NoStop}%
\bibitem [{\citenamefont {Bartkiewicz}\ \emph {et~al.}(2020)\citenamefont
  {Bartkiewicz}, \citenamefont {Gneiting}, \citenamefont {{\v{C}}ernoch},
  \citenamefont {Jir{\'a}kov{\'a}}, \citenamefont {Lemr},\ and\ \citenamefont
  {Nori}}]{bartkiewicz2020experimental}%
  \BibitemOpen
  \bibfield  {author} {\bibinfo {author} {\bibfnamefont {Karol}\ \bibnamefont
  {Bartkiewicz}}, \bibinfo {author} {\bibfnamefont {Clemens}\ \bibnamefont
  {Gneiting}}, \bibinfo {author} {\bibfnamefont {Anton{\'\i}n}\ \bibnamefont
  {{\v{C}}ernoch}}, \bibinfo {author} {\bibfnamefont {Kate{\v{r}}ina}\
  \bibnamefont {Jir{\'a}kov{\'a}}}, \bibinfo {author} {\bibfnamefont {Karel}\
  \bibnamefont {Lemr}}, \ and\ \bibinfo {author} {\bibfnamefont {Franco}\
  \bibnamefont {Nori}},\ }\bibfield  {title} {\enquote {\bibinfo {title}
  {Experimental kernel-based quantum machine learning in finite feature
  space},}\ }\href@noop {} {\bibfield  {journal} {\bibinfo  {journal}
  {Scientific Reports}\ }\textbf {\bibinfo {volume} {10}},\ \bibinfo {pages}
  {1--9} (\bibinfo {year} {2020})}\BibitemShut {NoStop}%
\bibitem [{\citenamefont {Blank}\ \emph {et~al.}(2020)\citenamefont {Blank},
  \citenamefont {Park}, \citenamefont {Rhee},\ and\ \citenamefont
  {Petruccione}}]{blank2020quantum}%
  \BibitemOpen
  \bibfield  {author} {\bibinfo {author} {\bibfnamefont {Carsten}\ \bibnamefont
  {Blank}}, \bibinfo {author} {\bibfnamefont {Daniel~K}\ \bibnamefont {Park}},
  \bibinfo {author} {\bibfnamefont {June-Koo~Kevin}\ \bibnamefont {Rhee}}, \
  and\ \bibinfo {author} {\bibfnamefont {Francesco}\ \bibnamefont
  {Petruccione}},\ }\bibfield  {title} {\enquote {\bibinfo {title} {Quantum
  classifier with tailored quantum kernel},}\ }\href@noop {} {\bibfield
  {journal} {\bibinfo  {journal} {npj Quantum Information}\ }\textbf {\bibinfo
  {volume} {6}},\ \bibinfo {pages} {1--7} (\bibinfo {year} {2020})}\BibitemShut
  {NoStop}%
\bibitem [{\citenamefont {Guan}\ \emph {et~al.}(2020)\citenamefont {Guan},
  \citenamefont {Perdue}, \citenamefont {Pesah}, \citenamefont {Schuld},
  \citenamefont {Terashi}, \citenamefont {Vallecorsa} \emph
  {et~al.}}]{guan2020quantum}%
  \BibitemOpen
  \bibfield  {author} {\bibinfo {author} {\bibfnamefont {Wen}\ \bibnamefont
  {Guan}}, \bibinfo {author} {\bibfnamefont {Gabriel}\ \bibnamefont {Perdue}},
  \bibinfo {author} {\bibfnamefont {Arthur}\ \bibnamefont {Pesah}}, \bibinfo
  {author} {\bibfnamefont {Maria}\ \bibnamefont {Schuld}}, \bibinfo {author}
  {\bibfnamefont {Koji}\ \bibnamefont {Terashi}}, \bibinfo {author}
  {\bibfnamefont {Sofia}\ \bibnamefont {Vallecorsa}},  \emph {et~al.},\
  }\bibfield  {title} {\enquote {\bibinfo {title} {Quantum machine learning in
  high energy physics},}\ }\href@noop {} {\bibfield  {journal} {\bibinfo
  {journal} {Machine Learning: Science and Technology}\ } (\bibinfo {year}
  {2020})}\BibitemShut {NoStop}%
\bibitem [{\citenamefont {Peters}\ \emph {et~al.}(2021)\citenamefont {Peters},
  \citenamefont {Caldeira}, \citenamefont {Ho}, \citenamefont {Leichenauer},
  \citenamefont {Mohseni}, \citenamefont {Neven}, \citenamefont {Spentzouris},
  \citenamefont {Strain},\ and\ \citenamefont {Perdue}}]{peters2021machine}%
  \BibitemOpen
  \bibfield  {author} {\bibinfo {author} {\bibfnamefont {Evan}\ \bibnamefont
  {Peters}}, \bibinfo {author} {\bibfnamefont {Jo{\~a}o}\ \bibnamefont
  {Caldeira}}, \bibinfo {author} {\bibfnamefont {Alan}\ \bibnamefont {Ho}},
  \bibinfo {author} {\bibfnamefont {Stefan}\ \bibnamefont {Leichenauer}},
  \bibinfo {author} {\bibfnamefont {Masoud}\ \bibnamefont {Mohseni}}, \bibinfo
  {author} {\bibfnamefont {Hartmut}\ \bibnamefont {Neven}}, \bibinfo {author}
  {\bibfnamefont {Panagiotis}\ \bibnamefont {Spentzouris}}, \bibinfo {author}
  {\bibfnamefont {Doug}\ \bibnamefont {Strain}}, \ and\ \bibinfo {author}
  {\bibfnamefont {Gabriel~N}\ \bibnamefont {Perdue}},\ }\bibfield  {title}
  {\enquote {\bibinfo {title} {Machine learning of high dimensional data on a
  noisy quantum processor},}\ }\href@noop {} {\bibfield  {journal} {\bibinfo
  {journal} {npj Quantum Information}\ }\textbf {\bibinfo {volume} {7}},\
  \bibinfo {pages} {1--5} (\bibinfo {year} {2021})}\BibitemShut {NoStop}%
\bibitem [{\citenamefont {Wu}\ \emph {et~al.}(2021{\natexlab{b}})\citenamefont
  {Wu}, \citenamefont {Sun}, \citenamefont {Guan}, \citenamefont {Zhou},
  \citenamefont {Chan}, \citenamefont {Cheng}, \citenamefont {Pham},
  \citenamefont {Qian}, \citenamefont {Wang}, \citenamefont {Zhang} \emph
  {et~al.}}]{wu2021application}%
  \BibitemOpen
  \bibfield  {author} {\bibinfo {author} {\bibfnamefont {Sau~Lan}\ \bibnamefont
  {Wu}}, \bibinfo {author} {\bibfnamefont {Shaojun}\ \bibnamefont {Sun}},
  \bibinfo {author} {\bibfnamefont {Wen}\ \bibnamefont {Guan}}, \bibinfo
  {author} {\bibfnamefont {Chen}\ \bibnamefont {Zhou}}, \bibinfo {author}
  {\bibfnamefont {Jay}\ \bibnamefont {Chan}}, \bibinfo {author} {\bibfnamefont
  {Chi~Lung}\ \bibnamefont {Cheng}}, \bibinfo {author} {\bibfnamefont {Tuan}\
  \bibnamefont {Pham}}, \bibinfo {author} {\bibfnamefont {Yan}\ \bibnamefont
  {Qian}}, \bibinfo {author} {\bibfnamefont {Alex~Zeng}\ \bibnamefont {Wang}},
  \bibinfo {author} {\bibfnamefont {Rui}\ \bibnamefont {Zhang}},  \emph
  {et~al.},\ }\bibfield  {title} {\enquote {\bibinfo {title} {Application of
  quantum machine learning using the quantum kernel algorithm on high energy
  physics analysis at the lhc},}\ }\href@noop {} {\bibfield  {journal}
  {\bibinfo  {journal} {Physical Review Research}\ }\textbf {\bibinfo {volume}
  {3}},\ \bibinfo {pages} {033221} (\bibinfo {year}
  {2021}{\natexlab{b}})}\BibitemShut {NoStop}%
\bibitem [{\citenamefont {Havl{\'\i}{\v{c}}ek}\ \emph
  {et~al.}(2019)\citenamefont {Havl{\'\i}{\v{c}}ek}, \citenamefont
  {C{\'o}rcoles}, \citenamefont {Temme}, \citenamefont {Harrow}, \citenamefont
  {Kandala}, \citenamefont {Chow},\ and\ \citenamefont
  {Gambetta}}]{havlivcek2019supervised}%
  \BibitemOpen
  \bibfield  {author} {\bibinfo {author} {\bibfnamefont {Vojt{\v{e}}ch}\
  \bibnamefont {Havl{\'\i}{\v{c}}ek}}, \bibinfo {author} {\bibfnamefont
  {Antonio~D}\ \bibnamefont {C{\'o}rcoles}}, \bibinfo {author} {\bibfnamefont
  {Kristan}\ \bibnamefont {Temme}}, \bibinfo {author} {\bibfnamefont {Aram~W}\
  \bibnamefont {Harrow}}, \bibinfo {author} {\bibfnamefont {Abhinav}\
  \bibnamefont {Kandala}}, \bibinfo {author} {\bibfnamefont {Jerry~M}\
  \bibnamefont {Chow}}, \ and\ \bibinfo {author} {\bibfnamefont {Jay~M}\
  \bibnamefont {Gambetta}},\ }\bibfield  {title} {\enquote {\bibinfo {title}
  {Supervised learning with quantum-enhanced feature spaces},}\ }\href@noop {}
  {\bibfield  {journal} {\bibinfo  {journal} {Nature}\ }\textbf {\bibinfo
  {volume} {567}},\ \bibinfo {pages} {209--212} (\bibinfo {year}
  {2019})}\BibitemShut {NoStop}%
\bibitem [{\citenamefont {Johri}\ \emph {et~al.}(2021)\citenamefont {Johri},
  \citenamefont {Debnath}, \citenamefont {Mocherla}, \citenamefont {Singk},
  \citenamefont {Prakash}, \citenamefont {Kim},\ and\ \citenamefont
  {Kerenidis}}]{johri2020nearest}%
  \BibitemOpen
  \bibfield  {author} {\bibinfo {author} {\bibfnamefont {Sonika}\ \bibnamefont
  {Johri}}, \bibinfo {author} {\bibfnamefont {Shantanu}\ \bibnamefont
  {Debnath}}, \bibinfo {author} {\bibfnamefont {Avinash}\ \bibnamefont
  {Mocherla}}, \bibinfo {author} {\bibfnamefont {Alexandros}\ \bibnamefont
  {Singk}}, \bibinfo {author} {\bibfnamefont {Anupam}\ \bibnamefont {Prakash}},
  \bibinfo {author} {\bibfnamefont {Jungsang}\ \bibnamefont {Kim}}, \ and\
  \bibinfo {author} {\bibfnamefont {Iordanis}\ \bibnamefont {Kerenidis}},\
  }\bibfield  {title} {\enquote {\bibinfo {title} {Nearest centroid
  classification on a trapped ion quantum computer},}\ }\href@noop {}
  {\bibfield  {journal} {\bibinfo  {journal} {npj Quantum Information}\
  }\textbf {\bibinfo {volume} {7}},\ \bibinfo {pages} {1--11} (\bibinfo {year}
  {2021})}\BibitemShut {NoStop}%
\bibitem [{\citenamefont {Huang}\ \emph
  {et~al.}(2021{\natexlab{b}})\citenamefont {Huang}, \citenamefont {Du},
  \citenamefont {Gong}, \citenamefont {Zhao}, \citenamefont {Wu}, \citenamefont
  {Wang}, \citenamefont {Li}, \citenamefont {Liang}, \citenamefont {Lin},
  \citenamefont {Xu} \emph {et~al.}}]{huang2020experimental}%
  \BibitemOpen
  \bibfield  {author} {\bibinfo {author} {\bibfnamefont {He-Liang}\
  \bibnamefont {Huang}}, \bibinfo {author} {\bibfnamefont {Yuxuan}\
  \bibnamefont {Du}}, \bibinfo {author} {\bibfnamefont {Ming}\ \bibnamefont
  {Gong}}, \bibinfo {author} {\bibfnamefont {Youwei}\ \bibnamefont {Zhao}},
  \bibinfo {author} {\bibfnamefont {Yulin}\ \bibnamefont {Wu}}, \bibinfo
  {author} {\bibfnamefont {Chaoyue}\ \bibnamefont {Wang}}, \bibinfo {author}
  {\bibfnamefont {Shaowei}\ \bibnamefont {Li}}, \bibinfo {author}
  {\bibfnamefont {Futian}\ \bibnamefont {Liang}}, \bibinfo {author}
  {\bibfnamefont {Jin}\ \bibnamefont {Lin}}, \bibinfo {author} {\bibfnamefont
  {Yu}~\bibnamefont {Xu}},  \emph {et~al.},\ }\bibfield  {title} {\enquote
  {\bibinfo {title} {Experimental quantum generative adversarial networks for
  image generation},}\ }\href@noop {} {\bibfield  {journal} {\bibinfo
  {journal} {Physical Review Applied}\ }\textbf {\bibinfo {volume} {16}},\
  \bibinfo {pages} {024051} (\bibinfo {year} {2021}{\natexlab{b}})}\BibitemShut
  {NoStop}%
\bibitem [{\citenamefont {Hubregtsen}\ \emph {et~al.}(2021)\citenamefont
  {Hubregtsen}, \citenamefont {Wierichs}, \citenamefont {Gil-Fuster},
  \citenamefont {Derks}, \citenamefont {Faehrmann},\ and\ \citenamefont
  {Meyer}}]{hubregtsen2021training}%
  \BibitemOpen
  \bibfield  {author} {\bibinfo {author} {\bibfnamefont {Thomas}\ \bibnamefont
  {Hubregtsen}}, \bibinfo {author} {\bibfnamefont {David}\ \bibnamefont
  {Wierichs}}, \bibinfo {author} {\bibfnamefont {Elies}\ \bibnamefont
  {Gil-Fuster}}, \bibinfo {author} {\bibfnamefont {Peter-Jan~HS}\ \bibnamefont
  {Derks}}, \bibinfo {author} {\bibfnamefont {Paul~K}\ \bibnamefont
  {Faehrmann}}, \ and\ \bibinfo {author} {\bibfnamefont {Johannes~Jakob}\
  \bibnamefont {Meyer}},\ }\bibfield  {title} {\enquote {\bibinfo {title}
  {Training quantum embedding kernels on near-term quantum computers},}\
  }\href@noop {} {\bibfield  {journal} {\bibinfo  {journal} {arXiv:2105.02276}\
  } (\bibinfo {year} {2021})}\BibitemShut {NoStop}%
\bibitem [{\citenamefont {Kusumoto}\ \emph {et~al.}(2021)\citenamefont
  {Kusumoto}, \citenamefont {Mitarai}, \citenamefont {Fujii}, \citenamefont
  {Kitagawa},\ and\ \citenamefont {Negoro}}]{kusumoto2021experimental}%
  \BibitemOpen
  \bibfield  {author} {\bibinfo {author} {\bibfnamefont {Takeru}\ \bibnamefont
  {Kusumoto}}, \bibinfo {author} {\bibfnamefont {Kosuke}\ \bibnamefont
  {Mitarai}}, \bibinfo {author} {\bibfnamefont {Keisuke}\ \bibnamefont
  {Fujii}}, \bibinfo {author} {\bibfnamefont {Masahiro}\ \bibnamefont
  {Kitagawa}}, \ and\ \bibinfo {author} {\bibfnamefont {Makoto}\ \bibnamefont
  {Negoro}},\ }\bibfield  {title} {\enquote {\bibinfo {title} {Experimental
  quantum kernel trick with nuclear spins in a solid},}\ }\href@noop {}
  {\bibfield  {journal} {\bibinfo  {journal} {npj Quantum Information}\
  }\textbf {\bibinfo {volume} {7}},\ \bibinfo {pages} {1--7} (\bibinfo {year}
  {2021})}\BibitemShut {NoStop}%
\bibitem [{\citenamefont {Dutta}\ \emph {et~al.}(2021)\citenamefont {Dutta},
  \citenamefont {P{\'e}rez-Salinas}, \citenamefont {Cheng}, \citenamefont
  {Latorre},\ and\ \citenamefont {Mukherjee}}]{dutta2021realization}%
  \BibitemOpen
  \bibfield  {author} {\bibinfo {author} {\bibfnamefont {Tarun}\ \bibnamefont
  {Dutta}}, \bibinfo {author} {\bibfnamefont {Adri{\'a}n}\ \bibnamefont
  {P{\'e}rez-Salinas}}, \bibinfo {author} {\bibfnamefont {Jasper Phua~Sing}\
  \bibnamefont {Cheng}}, \bibinfo {author} {\bibfnamefont {Jos{\'e}~Ignacio}\
  \bibnamefont {Latorre}}, \ and\ \bibinfo {author} {\bibfnamefont {Manas}\
  \bibnamefont {Mukherjee}},\ }\bibfield  {title} {\enquote {\bibinfo {title}
  {Realization of an ion trap quantum classifier},}\ }\href@noop {} {\bibfield
  {journal} {\bibinfo  {journal} {arXiv preprint arXiv:2106.14059}\ } (\bibinfo
  {year} {2021})}\BibitemShut {NoStop}%
\bibitem [{\citenamefont {P{\'e}rez-Salinas}\ \emph {et~al.}(2020)\citenamefont
  {P{\'e}rez-Salinas}, \citenamefont {Cervera-Lierta}, \citenamefont
  {Gil-Fuster},\ and\ \citenamefont {Latorre}}]{perez2020data}%
  \BibitemOpen
  \bibfield  {author} {\bibinfo {author} {\bibfnamefont {Adri{\'a}n}\
  \bibnamefont {P{\'e}rez-Salinas}}, \bibinfo {author} {\bibfnamefont {Alba}\
  \bibnamefont {Cervera-Lierta}}, \bibinfo {author} {\bibfnamefont {Elies}\
  \bibnamefont {Gil-Fuster}}, \ and\ \bibinfo {author} {\bibfnamefont
  {Jos{\'e}~I}\ \bibnamefont {Latorre}},\ }\bibfield  {title} {\enquote
  {\bibinfo {title} {Data re-uploading for a universal quantum classifier},}\
  }\href@noop {} {\bibfield  {journal} {\bibinfo  {journal} {Quantum}\ }\textbf
  {\bibinfo {volume} {4}},\ \bibinfo {pages} {226} (\bibinfo {year}
  {2020})}\BibitemShut {NoStop}%
\bibitem [{\citenamefont {Schuld}(2021)}]{schuld2021quantum}%
  \BibitemOpen
  \bibfield  {author} {\bibinfo {author} {\bibfnamefont {Maria}\ \bibnamefont
  {Schuld}},\ }\bibfield  {title} {\enquote {\bibinfo {title} {Quantum machine
  learning models are kernel methods},}\ }\href@noop {} {\bibfield  {journal}
  {\bibinfo  {journal} {arXiv:2101.11020}\ } (\bibinfo {year}
  {2021})}\BibitemShut {NoStop}%
\bibitem [{\citenamefont {Temme}\ \emph {et~al.}(2017)\citenamefont {Temme},
  \citenamefont {Bravyi},\ and\ \citenamefont {Gambetta}}]{temme2017error}%
  \BibitemOpen
  \bibfield  {author} {\bibinfo {author} {\bibfnamefont {Kristan}\ \bibnamefont
  {Temme}}, \bibinfo {author} {\bibfnamefont {Sergey}\ \bibnamefont {Bravyi}},
  \ and\ \bibinfo {author} {\bibfnamefont {Jay~M}\ \bibnamefont {Gambetta}},\
  }\bibfield  {title} {\enquote {\bibinfo {title} {Error mitigation for
  short-depth quantum circuits},}\ }\href@noop {} {\bibfield  {journal}
  {\bibinfo  {journal} {Physical review letters}\ }\textbf {\bibinfo {volume}
  {119}},\ \bibinfo {pages} {180509} (\bibinfo {year} {2017})}\BibitemShut
  {NoStop}%
\bibitem [{\citenamefont {Endo}\ \emph {et~al.}(2021)\citenamefont {Endo},
  \citenamefont {Cai}, \citenamefont {Benjamin},\ and\ \citenamefont
  {Yuan}}]{endo2021hybrid}%
  \BibitemOpen
  \bibfield  {author} {\bibinfo {author} {\bibfnamefont {Suguru}\ \bibnamefont
  {Endo}}, \bibinfo {author} {\bibfnamefont {Zhenyu}\ \bibnamefont {Cai}},
  \bibinfo {author} {\bibfnamefont {Simon~C}\ \bibnamefont {Benjamin}}, \ and\
  \bibinfo {author} {\bibfnamefont {Xiao}\ \bibnamefont {Yuan}},\ }\bibfield
  {title} {\enquote {\bibinfo {title} {Hybrid quantum-classical algorithms and
  quantum error mitigation},}\ }\href@noop {} {\bibfield  {journal} {\bibinfo
  {journal} {Journal of the Physical Society of Japan}\ }\textbf {\bibinfo
  {volume} {90}},\ \bibinfo {pages} {032001} (\bibinfo {year}
  {2021})}\BibitemShut {NoStop}%
\bibitem [{\citenamefont {Scholkopf}\ and\ \citenamefont
  {Smola}(2018)}]{scholkopf2018learning}%
  \BibitemOpen
  \bibfield  {author} {\bibinfo {author} {\bibfnamefont {Bernhard}\
  \bibnamefont {Scholkopf}}\ and\ \bibinfo {author} {\bibfnamefont
  {Alexander~J}\ \bibnamefont {Smola}},\ }\href@noop {} {\emph {\bibinfo
  {title} {Learning with kernels: support vector machines, regularization,
  optimization, and beyond}}}\ (\bibinfo  {publisher} {Adaptive Computation and
  Machine Learning Series},\ \bibinfo {year} {2018})\BibitemShut {NoStop}%
\bibitem [{\citenamefont {Wolkowicz}\ \emph {et~al.}(2012)\citenamefont
  {Wolkowicz}, \citenamefont {Saigal},\ and\ \citenamefont
  {Vandenberghe}}]{wolkowicz2012handbook}%
  \BibitemOpen
  \bibfield  {author} {\bibinfo {author} {\bibfnamefont {Henry}\ \bibnamefont
  {Wolkowicz}}, \bibinfo {author} {\bibfnamefont {Romesh}\ \bibnamefont
  {Saigal}}, \ and\ \bibinfo {author} {\bibfnamefont {Lieven}\ \bibnamefont
  {Vandenberghe}},\ }\href@noop {} {\emph {\bibinfo {title} {Handbook of
  semidefinite programming: theory, algorithms, and applications}}},\
  Vol.~\bibinfo {volume} {27}\ (\bibinfo  {publisher} {Springer Science \&
  Business Media},\ \bibinfo {year} {2012})\BibitemShut {NoStop}%
\bibitem [{\citenamefont {Brand{\~a}o}\ \emph {et~al.}(2017)\citenamefont
  {Brand{\~a}o}, \citenamefont {Kalev}, \citenamefont {Li}, \citenamefont
  {Lin}, \citenamefont {Svore},\ and\ \citenamefont {Wu}}]{brandao2017quantum}%
  \BibitemOpen
  \bibfield  {author} {\bibinfo {author} {\bibfnamefont {Fernando~GSL}\
  \bibnamefont {Brand{\~a}o}}, \bibinfo {author} {\bibfnamefont {Amir}\
  \bibnamefont {Kalev}}, \bibinfo {author} {\bibfnamefont {Tongyang}\
  \bibnamefont {Li}}, \bibinfo {author} {\bibfnamefont {Cedric Yen-Yu}\
  \bibnamefont {Lin}}, \bibinfo {author} {\bibfnamefont {Krysta~M}\
  \bibnamefont {Svore}}, \ and\ \bibinfo {author} {\bibfnamefont {Xiaodi}\
  \bibnamefont {Wu}},\ }\bibfield  {title} {\enquote {\bibinfo {title} {Quantum
  sdp solvers: Large speed-ups, optimality, and applications to quantum
  learning},}\ }\href@noop {} {\bibfield  {journal} {\bibinfo  {journal}
  {arXiv:1710.02581}\ } (\bibinfo {year} {2017})}\BibitemShut {NoStop}%
\bibitem [{\citenamefont {Bharti}\ \emph
  {et~al.}(2022{\natexlab{b}})\citenamefont {Bharti}, \citenamefont {Haug},
  \citenamefont {Vedral},\ and\ \citenamefont {Kwek}}]{bharti2021nisq}%
  \BibitemOpen
  \bibfield  {author} {\bibinfo {author} {\bibfnamefont {Kishor}\ \bibnamefont
  {Bharti}}, \bibinfo {author} {\bibfnamefont {Tobias}\ \bibnamefont {Haug}},
  \bibinfo {author} {\bibfnamefont {Vlatko}\ \bibnamefont {Vedral}}, \ and\
  \bibinfo {author} {\bibfnamefont {Leong-Chuan}\ \bibnamefont {Kwek}},\
  }\bibfield  {title} {\enquote {\bibinfo {title} {Noisy intermediate-scale
  quantum algorithm for semidefinite programming},}\ }\href@noop {} {\bibfield
  {journal} {\bibinfo  {journal} {Phys. Rev. A}\ }\textbf {\bibinfo {volume}
  {105}},\ \bibinfo {pages} {052445} (\bibinfo {year}
  {2022}{\natexlab{b}})}\BibitemShut {NoStop}%
\bibitem [{\citenamefont {Kandala}\ \emph {et~al.}(2017)\citenamefont
  {Kandala}, \citenamefont {Mezzacapo}, \citenamefont {Temme}, \citenamefont
  {Takita}, \citenamefont {Brink}, \citenamefont {Chow},\ and\ \citenamefont
  {Gambetta}}]{kandala2017hardware}%
  \BibitemOpen
  \bibfield  {author} {\bibinfo {author} {\bibfnamefont {Abhinav}\ \bibnamefont
  {Kandala}}, \bibinfo {author} {\bibfnamefont {Antonio}\ \bibnamefont
  {Mezzacapo}}, \bibinfo {author} {\bibfnamefont {Kristan}\ \bibnamefont
  {Temme}}, \bibinfo {author} {\bibfnamefont {Maika}\ \bibnamefont {Takita}},
  \bibinfo {author} {\bibfnamefont {Markus}\ \bibnamefont {Brink}}, \bibinfo
  {author} {\bibfnamefont {Jerry~M}\ \bibnamefont {Chow}}, \ and\ \bibinfo
  {author} {\bibfnamefont {Jay~M}\ \bibnamefont {Gambetta}},\ }\bibfield
  {title} {\enquote {\bibinfo {title} {Hardware-efficient variational quantum
  eigensolver for small molecules and quantum magnets},}\ }\href@noop {}
  {\bibfield  {journal} {\bibinfo  {journal} {Nature}\ }\textbf {\bibinfo
  {volume} {549}},\ \bibinfo {pages} {242} (\bibinfo {year}
  {2017})}\BibitemShut {NoStop}%
\bibitem [{\citenamefont {Haug}\ \emph {et~al.}(2021)\citenamefont {Haug},
  \citenamefont {Bharti},\ and\ \citenamefont {Kim}}]{haug2021capacity}%
  \BibitemOpen
  \bibfield  {author} {\bibinfo {author} {\bibfnamefont {Tobias}\ \bibnamefont
  {Haug}}, \bibinfo {author} {\bibfnamefont {Kishor}\ \bibnamefont {Bharti}}, \
  and\ \bibinfo {author} {\bibfnamefont {M.S.}\ \bibnamefont {Kim}},\
  }\bibfield  {title} {\enquote {\bibinfo {title} {Capacity and quantum
  geometry of parametrized quantum circuits},}\ }\href {\doibase
  10.1103/PRXQuantum.2.040309} {\bibfield  {journal} {\bibinfo  {journal} {PRX
  Quantum}\ }\textbf {\bibinfo {volume} {2}},\ \bibinfo {pages} {040309}
  (\bibinfo {year} {2021})}\BibitemShut {NoStop}%
\bibitem [{\citenamefont {Meyer}(2021)}]{meyer2021fisher}%
  \BibitemOpen
  \bibfield  {author} {\bibinfo {author} {\bibfnamefont {Johannes~Jakob}\
  \bibnamefont {Meyer}},\ }\bibfield  {title} {\enquote {\bibinfo {title}
  {Fisher information in noisy intermediate-scale quantum applications},}\
  }\href@noop {} {\bibfield  {journal} {\bibinfo  {journal} {Quantum}\ }\textbf
  {\bibinfo {volume} {5}},\ \bibinfo {pages} {539} (\bibinfo {year}
  {2021})}\BibitemShut {NoStop}%
\bibitem [{\citenamefont {Banchi}\ \emph {et~al.}(2021)\citenamefont {Banchi},
  \citenamefont {Pereira},\ and\ \citenamefont
  {Pirandola}}]{banchi2021generalization}%
  \BibitemOpen
  \bibfield  {author} {\bibinfo {author} {\bibfnamefont {Leonardo}\
  \bibnamefont {Banchi}}, \bibinfo {author} {\bibfnamefont {Jason}\
  \bibnamefont {Pereira}}, \ and\ \bibinfo {author} {\bibfnamefont {Stefano}\
  \bibnamefont {Pirandola}},\ }\bibfield  {title} {\enquote {\bibinfo {title}
  {Generalization in quantum machine learning: A quantum information
  standpoint},}\ }\href {\doibase 10.1103/PRXQuantum.2.040321} {\bibfield
  {journal} {\bibinfo  {journal} {PRX Quantum}\ }\textbf {\bibinfo {volume}
  {2}},\ \bibinfo {pages} {040321} (\bibinfo {year} {2021})}\BibitemShut
  {NoStop}%
\bibitem [{\citenamefont {Haug}\ and\ \citenamefont
  {Kim}(2021)}]{haug2021optimal}%
  \BibitemOpen
  \bibfield  {author} {\bibinfo {author} {\bibfnamefont {Tobias}\ \bibnamefont
  {Haug}}\ and\ \bibinfo {author} {\bibfnamefont {M.~S.}\ \bibnamefont {Kim}},\
  }\bibfield  {title} {\enquote {\bibinfo {title} {Optimal training of
  variational quantum algorithms without barren plateaus},}\ }\href@noop {}
  {\bibfield  {journal} {\bibinfo  {journal} {arXiv:2104.14543}\ } (\bibinfo
  {year} {2021})}\BibitemShut {NoStop}%
\bibitem [{\citenamefont {Goodfellow}\ \emph {et~al.}(2016)\citenamefont
  {Goodfellow}, \citenamefont {Bengio},\ and\ \citenamefont
  {Courville}}]{goodfellow2016deep}%
  \BibitemOpen
  \bibfield  {author} {\bibinfo {author} {\bibfnamefont {Ian}\ \bibnamefont
  {Goodfellow}}, \bibinfo {author} {\bibfnamefont {Yoshua}\ \bibnamefont
  {Bengio}}, \ and\ \bibinfo {author} {\bibfnamefont {Aaron}\ \bibnamefont
  {Courville}},\ }\href@noop {} {\emph {\bibinfo {title} {Deep learning}}}\
  (\bibinfo  {publisher} {MIT press},\ \bibinfo {year} {2016})\BibitemShut
  {NoStop}%
\bibitem [{\citenamefont {Haug}\ and\ \citenamefont
  {Kim}(2022)}]{haug2021natural}%
  \BibitemOpen
  \bibfield  {author} {\bibinfo {author} {\bibfnamefont {Tobias}\ \bibnamefont
  {Haug}}\ and\ \bibinfo {author} {\bibfnamefont {M.~S.}\ \bibnamefont {Kim}},\
  }\bibfield  {title} {\enquote {\bibinfo {title} {Natural parametrized quantum
  circuit},}\ }\href {\doibase 10.1103/PhysRevA.106.052611} {\bibfield
  {journal} {\bibinfo  {journal} {Phys. Rev. A}\ }\textbf {\bibinfo {volume}
  {106}},\ \bibinfo {pages} {052611} (\bibinfo {year} {2022})}\BibitemShut
  {NoStop}%
\bibitem [{\citenamefont {Elben}\ \emph {et~al.}(2019)\citenamefont {Elben},
  \citenamefont {Vermersch}, \citenamefont {Roos},\ and\ \citenamefont
  {Zoller}}]{elben2019statistical}%
  \BibitemOpen
  \bibfield  {author} {\bibinfo {author} {\bibfnamefont {Andreas}\ \bibnamefont
  {Elben}}, \bibinfo {author} {\bibfnamefont {Beno{\^\i}t}\ \bibnamefont
  {Vermersch}}, \bibinfo {author} {\bibfnamefont {Christian~F}\ \bibnamefont
  {Roos}}, \ and\ \bibinfo {author} {\bibfnamefont {Peter}\ \bibnamefont
  {Zoller}},\ }\bibfield  {title} {\enquote {\bibinfo {title} {Statistical
  correlations between locally randomized measurements: A toolbox for probing
  entanglement in many-body quantum states},}\ }\href@noop {} {\bibfield
  {journal} {\bibinfo  {journal} {Physical Review A}\ }\textbf {\bibinfo
  {volume} {99}},\ \bibinfo {pages} {052323} (\bibinfo {year}
  {2019})}\BibitemShut {NoStop}%
\bibitem [{\citenamefont {Elben}\ \emph {et~al.}(2020)\citenamefont {Elben},
  \citenamefont {Vermersch}, \citenamefont {van Bijnen}, \citenamefont
  {Kokail}, \citenamefont {Brydges}, \citenamefont {Maier}, \citenamefont
  {Joshi}, \citenamefont {Blatt}, \citenamefont {Roos},\ and\ \citenamefont
  {Zoller}}]{elben2020cross}%
  \BibitemOpen
  \bibfield  {author} {\bibinfo {author} {\bibfnamefont {Andreas}\ \bibnamefont
  {Elben}}, \bibinfo {author} {\bibfnamefont {Beno{\^\i}t}\ \bibnamefont
  {Vermersch}}, \bibinfo {author} {\bibfnamefont {Rick}\ \bibnamefont {van
  Bijnen}}, \bibinfo {author} {\bibfnamefont {Christian}\ \bibnamefont
  {Kokail}}, \bibinfo {author} {\bibfnamefont {Tiff}\ \bibnamefont {Brydges}},
  \bibinfo {author} {\bibfnamefont {Christine}\ \bibnamefont {Maier}}, \bibinfo
  {author} {\bibfnamefont {Manoj~K}\ \bibnamefont {Joshi}}, \bibinfo {author}
  {\bibfnamefont {Rainer}\ \bibnamefont {Blatt}}, \bibinfo {author}
  {\bibfnamefont {Christian~F}\ \bibnamefont {Roos}}, \ and\ \bibinfo {author}
  {\bibfnamefont {Peter}\ \bibnamefont {Zoller}},\ }\bibfield  {title}
  {\enquote {\bibinfo {title} {Cross-platform verification of intermediate
  scale quantum devices},}\ }\href@noop {} {\bibfield  {journal} {\bibinfo
  {journal} {Physical review letters}\ }\textbf {\bibinfo {volume} {124}},\
  \bibinfo {pages} {010504} (\bibinfo {year} {2020})}\BibitemShut {NoStop}%
\bibitem [{\citenamefont {Zhu}\ \emph {et~al.}(2022)\citenamefont {Zhu},
  \citenamefont {Cian}, \citenamefont {Noel}, \citenamefont {Risinger},
  \citenamefont {Biswas}, \citenamefont {Egan}, \citenamefont {Zhu},
  \citenamefont {Green}, \citenamefont {Alderete}, \citenamefont {Nguyen} \emph
  {et~al.}}]{zhu2021crossplatform}%
  \BibitemOpen
  \bibfield  {author} {\bibinfo {author} {\bibfnamefont {D}~\bibnamefont
  {Zhu}}, \bibinfo {author} {\bibfnamefont {ZP}~\bibnamefont {Cian}}, \bibinfo
  {author} {\bibfnamefont {C}~\bibnamefont {Noel}}, \bibinfo {author}
  {\bibfnamefont {A}~\bibnamefont {Risinger}}, \bibinfo {author} {\bibfnamefont
  {D}~\bibnamefont {Biswas}}, \bibinfo {author} {\bibfnamefont {L}~\bibnamefont
  {Egan}}, \bibinfo {author} {\bibfnamefont {Y}~\bibnamefont {Zhu}}, \bibinfo
  {author} {\bibfnamefont {AM}~\bibnamefont {Green}}, \bibinfo {author}
  {\bibfnamefont {C~Huerta}\ \bibnamefont {Alderete}}, \bibinfo {author}
  {\bibfnamefont {NH}~\bibnamefont {Nguyen}},  \emph {et~al.},\ }\bibfield
  {title} {\enquote {\bibinfo {title} {Cross-platform comparison of arbitrary
  quantum states},}\ }\href@noop {} {\bibfield  {journal} {\bibinfo  {journal}
  {Nature communications}\ }\textbf {\bibinfo {volume} {13}},\ \bibinfo {pages}
  {1--6} (\bibinfo {year} {2022})}\BibitemShut {NoStop}%
\bibitem [{\citenamefont {Rath}\ \emph {et~al.}(2021)\citenamefont {Rath},
  \citenamefont {van Bijnen}, \citenamefont {Elben}, \citenamefont {Zoller},\
  and\ \citenamefont {Vermersch}}]{rath2021importance}%
  \BibitemOpen
  \bibfield  {author} {\bibinfo {author} {\bibfnamefont {Aniket}\ \bibnamefont
  {Rath}}, \bibinfo {author} {\bibfnamefont {Rick}\ \bibnamefont {van Bijnen}},
  \bibinfo {author} {\bibfnamefont {Andreas}\ \bibnamefont {Elben}}, \bibinfo
  {author} {\bibfnamefont {Peter}\ \bibnamefont {Zoller}}, \ and\ \bibinfo
  {author} {\bibfnamefont {Beno{\^\i}t}\ \bibnamefont {Vermersch}},\ }\bibfield
   {title} {\enquote {\bibinfo {title} {Importance sampling of randomized
  measurements for probing entanglement},}\ }\href@noop {} {\bibfield
  {journal} {\bibinfo  {journal} {Physical review letters}\ }\textbf {\bibinfo
  {volume} {127}},\ \bibinfo {pages} {200503} (\bibinfo {year}
  {2021})}\BibitemShut {NoStop}%
\bibitem [{\citenamefont {Buhrman}\ \emph {et~al.}(2001)\citenamefont
  {Buhrman}, \citenamefont {Cleve}, \citenamefont {Watrous},\ and\
  \citenamefont {De~Wolf}}]{buhrman2001quantum}%
  \BibitemOpen
  \bibfield  {author} {\bibinfo {author} {\bibfnamefont {Harry}\ \bibnamefont
  {Buhrman}}, \bibinfo {author} {\bibfnamefont {Richard}\ \bibnamefont
  {Cleve}}, \bibinfo {author} {\bibfnamefont {John}\ \bibnamefont {Watrous}}, \
  and\ \bibinfo {author} {\bibfnamefont {Ronald}\ \bibnamefont {De~Wolf}},\
  }\bibfield  {title} {\enquote {\bibinfo {title} {Quantum fingerprinting},}\
  }\href@noop {} {\bibfield  {journal} {\bibinfo  {journal} {Physical Review
  Letters}\ }\textbf {\bibinfo {volume} {87}},\ \bibinfo {pages} {167902}
  (\bibinfo {year} {2001})}\BibitemShut {NoStop}%
\bibitem [{\citenamefont {Nguyen}\ \emph {et~al.}(2021)\citenamefont {Nguyen},
  \citenamefont {Tseng}, \citenamefont {Maslennikov}, \citenamefont {Gan},\
  and\ \citenamefont {Matsukevich}}]{nguyen2021experimental}%
  \BibitemOpen
  \bibfield  {author} {\bibinfo {author} {\bibfnamefont {Chi-Huan}\
  \bibnamefont {Nguyen}}, \bibinfo {author} {\bibfnamefont {Ko-Wei}\
  \bibnamefont {Tseng}}, \bibinfo {author} {\bibfnamefont {Gleb}\ \bibnamefont
  {Maslennikov}}, \bibinfo {author} {\bibfnamefont {HCJ}\ \bibnamefont {Gan}},
  \ and\ \bibinfo {author} {\bibfnamefont {Dzmitry}\ \bibnamefont
  {Matsukevich}},\ }\bibfield  {title} {\enquote {\bibinfo {title}
  {Experimental swap test of infinite dimensional quantum states},}\
  }\href@noop {} {\bibfield  {journal} {\bibinfo  {journal} {arXiv preprint
  arXiv:2103.10219}\ } (\bibinfo {year} {2021})}\BibitemShut {NoStop}%
\bibitem [{\citenamefont {Vovrosh}\ \emph {et~al.}(2021)\citenamefont
  {Vovrosh}, \citenamefont {Khosla}, \citenamefont {Greenaway}, \citenamefont
  {Self}, \citenamefont {Kim},\ and\ \citenamefont
  {Knolle}}]{vovrosh2021efficient}%
  \BibitemOpen
  \bibfield  {author} {\bibinfo {author} {\bibfnamefont {Joseph}\ \bibnamefont
  {Vovrosh}}, \bibinfo {author} {\bibfnamefont {Kiran~E}\ \bibnamefont
  {Khosla}}, \bibinfo {author} {\bibfnamefont {Sean}\ \bibnamefont
  {Greenaway}}, \bibinfo {author} {\bibfnamefont {Christopher}\ \bibnamefont
  {Self}}, \bibinfo {author} {\bibfnamefont {Myungshik~S}\ \bibnamefont {Kim}},
  \ and\ \bibinfo {author} {\bibfnamefont {Johannes}\ \bibnamefont {Knolle}},\
  }\bibfield  {title} {\enquote {\bibinfo {title} {Simple mitigation of global
  depolarizing errors in quantum simulations},}\ }\href@noop {} {\bibfield
  {journal} {\bibinfo  {journal} {Physical Review E}\ }\textbf {\bibinfo
  {volume} {104}},\ \bibinfo {pages} {035309} (\bibinfo {year}
  {2021})}\BibitemShut {NoStop}%
\bibitem [{\citenamefont {Johansson}\ \emph {et~al.}(2012)\citenamefont
  {Johansson}, \citenamefont {Nation},\ and\ \citenamefont
  {Nori}}]{johansson2012qutip}%
  \BibitemOpen
  \bibfield  {author} {\bibinfo {author} {\bibfnamefont {J~Robert}\
  \bibnamefont {Johansson}}, \bibinfo {author} {\bibfnamefont {Paul~D}\
  \bibnamefont {Nation}}, \ and\ \bibinfo {author} {\bibfnamefont {Franco}\
  \bibnamefont {Nori}},\ }\bibfield  {title} {\enquote {\bibinfo {title}
  {Qutip: An open-source python framework for the dynamics of open quantum
  systems},}\ }\href@noop {} {\bibfield  {journal} {\bibinfo  {journal}
  {Computer Physics Communications}\ }\textbf {\bibinfo {volume} {183}},\
  \bibinfo {pages} {1760--1772} (\bibinfo {year} {2012})}\BibitemShut {NoStop}%
\bibitem [{\citenamefont {Luo}\ \emph {et~al.}(2020)\citenamefont {Luo},
  \citenamefont {Liu}, \citenamefont {Zhang},\ and\ \citenamefont
  {Wang}}]{yao}%
  \BibitemOpen
  \bibfield  {author} {\bibinfo {author} {\bibfnamefont {Xiu-Zhe}\ \bibnamefont
  {Luo}}, \bibinfo {author} {\bibfnamefont {Jin-Guo}\ \bibnamefont {Liu}},
  \bibinfo {author} {\bibfnamefont {Pan}\ \bibnamefont {Zhang}}, \ and\
  \bibinfo {author} {\bibfnamefont {Lei}\ \bibnamefont {Wang}},\ }\bibfield
  {title} {\enquote {\bibinfo {title} {Yao. jl: Extensible, efficient framework
  for quantum algorithm design},}\ }\href
  {https://doi.org/10.22331/q-2020-10-11-341} {\bibfield  {journal} {\bibinfo
  {journal} {Quantum}\ }\textbf {\bibinfo {volume} {4}},\ \bibinfo {pages}
  {341} (\bibinfo {year} {2020})}\BibitemShut {NoStop}%
\bibitem [{\citenamefont {McClean}\ \emph {et~al.}(2018)\citenamefont
  {McClean}, \citenamefont {Boixo}, \citenamefont {Smelyanskiy}, \citenamefont
  {Babbush},\ and\ \citenamefont {Neven}}]{mcclean2018barren}%
  \BibitemOpen
  \bibfield  {author} {\bibinfo {author} {\bibfnamefont {Jarrod~R}\
  \bibnamefont {McClean}}, \bibinfo {author} {\bibfnamefont {Sergio}\
  \bibnamefont {Boixo}}, \bibinfo {author} {\bibfnamefont {Vadim~N}\
  \bibnamefont {Smelyanskiy}}, \bibinfo {author} {\bibfnamefont {Ryan}\
  \bibnamefont {Babbush}}, \ and\ \bibinfo {author} {\bibfnamefont {Hartmut}\
  \bibnamefont {Neven}},\ }\bibfield  {title} {\enquote {\bibinfo {title}
  {Barren plateaus in quantum neural network training landscapes},}\
  }\href@noop {} {\bibfield  {journal} {\bibinfo  {journal} {Nature
  communications}\ }\textbf {\bibinfo {volume} {9}},\ \bibinfo {pages} {4812}
  (\bibinfo {year} {2018})}\BibitemShut {NoStop}%
\bibitem [{ibm()}]{ibmq-devices}%
  \BibitemOpen
  \href@noop {} {}\bibinfo {note} {\emph{ibmq\_guadalupe} (v1.3.4),
  \emph{ibmq\_toronto} (v1.5.7) IBM Quantum team. Retrieved from
  \url{https://quantum-computing.ibm.com} (2021).}\BibitemShut {Stop}%
\bibitem [{\citenamefont {Kaynak}(1995)}]{kaynak1995methods}%
  \BibitemOpen
  \bibfield  {author} {\bibinfo {author} {\bibfnamefont {C}~\bibnamefont
  {Kaynak}},\ }\bibfield  {title} {\enquote {\bibinfo {title} {Methods of
  combining multiple classifiers and their applications to handwritten digit
  recognition},}\ }\href@noop {} {\bibfield  {journal} {\bibinfo  {journal}
  {Unpublished master thesis, Bogazici University}\ } (\bibinfo {year}
  {1995})}\BibitemShut {NoStop}%
\bibitem [{\citenamefont {Pedregosa}\ \emph {et~al.}(2011)\citenamefont
  {Pedregosa}, \citenamefont {Varoquaux}, \citenamefont {Gramfort},
  \citenamefont {Michel}, \citenamefont {Thirion}, \citenamefont {Grisel},
  \citenamefont {Blondel}, \citenamefont {Prettenhofer}, \citenamefont {Weiss},
  \citenamefont {Dubourg}, \citenamefont {Vanderplas}, \citenamefont {Passos},
  \citenamefont {Cournapeau}, \citenamefont {Brucher}, \citenamefont {Perrot},\
  and\ \citenamefont {Duchesnay}}]{pedregosa2011scikit}%
  \BibitemOpen
  \bibfield  {author} {\bibinfo {author} {\bibfnamefont {F.}~\bibnamefont
  {Pedregosa}}, \bibinfo {author} {\bibfnamefont {G.}~\bibnamefont
  {Varoquaux}}, \bibinfo {author} {\bibfnamefont {A.}~\bibnamefont {Gramfort}},
  \bibinfo {author} {\bibfnamefont {V.}~\bibnamefont {Michel}}, \bibinfo
  {author} {\bibfnamefont {B.}~\bibnamefont {Thirion}}, \bibinfo {author}
  {\bibfnamefont {O.}~\bibnamefont {Grisel}}, \bibinfo {author} {\bibfnamefont
  {M.}~\bibnamefont {Blondel}}, \bibinfo {author} {\bibfnamefont
  {P.}~\bibnamefont {Prettenhofer}}, \bibinfo {author} {\bibfnamefont
  {R.}~\bibnamefont {Weiss}}, \bibinfo {author} {\bibfnamefont
  {V.}~\bibnamefont {Dubourg}}, \bibinfo {author} {\bibfnamefont
  {J.}~\bibnamefont {Vanderplas}}, \bibinfo {author} {\bibfnamefont
  {A.}~\bibnamefont {Passos}}, \bibinfo {author} {\bibfnamefont
  {D.}~\bibnamefont {Cournapeau}}, \bibinfo {author} {\bibfnamefont
  {M.}~\bibnamefont {Brucher}}, \bibinfo {author} {\bibfnamefont
  {M.}~\bibnamefont {Perrot}}, \ and\ \bibinfo {author} {\bibfnamefont
  {E.}~\bibnamefont {Duchesnay}},\ }\bibfield  {title} {\enquote {\bibinfo
  {title} {Scikit-learn: Machine learning in {P}ython},}\ }\href@noop {}
  {\bibfield  {journal} {\bibinfo  {journal} {Journal of Machine Learning
  Research}\ }\textbf {\bibinfo {volume} {12}},\ \bibinfo {pages} {2825--2830}
  (\bibinfo {year} {2011})}\BibitemShut {NoStop}%
\bibitem [{\citenamefont {Schreiber}\ \emph {et~al.}(2022)\citenamefont
  {Schreiber}, \citenamefont {Eisert},\ and\ \citenamefont
  {Meyer}}]{schreiber2022classical}%
  \BibitemOpen
  \bibfield  {author} {\bibinfo {author} {\bibfnamefont {Franz~J}\ \bibnamefont
  {Schreiber}}, \bibinfo {author} {\bibfnamefont {Jens}\ \bibnamefont
  {Eisert}}, \ and\ \bibinfo {author} {\bibfnamefont {Johannes~Jakob}\
  \bibnamefont {Meyer}},\ }\bibfield  {title} {\enquote {\bibinfo {title}
  {Classical surrogates for quantum learning models},}\ }\href@noop {}
  {\bibfield  {journal} {\bibinfo  {journal} {arXiv:2206.11740}\ } (\bibinfo
  {year} {2022})}\BibitemShut {NoStop}%
\bibitem [{\citenamefont {Chatterjee}\ and\ \citenamefont
  {Yu}(2016)}]{chatterjee2016generalized}%
  \BibitemOpen
  \bibfield  {author} {\bibinfo {author} {\bibfnamefont {Rupak}\ \bibnamefont
  {Chatterjee}}\ and\ \bibinfo {author} {\bibfnamefont {Ting}\ \bibnamefont
  {Yu}},\ }\bibfield  {title} {\enquote {\bibinfo {title} {Generalized coherent
  states, reproducing kernels, and quantum support vector machines},}\
  }\href@noop {} {\bibfield  {journal} {\bibinfo  {journal} {arXiv:1612.03713}\
  } (\bibinfo {year} {2016})}\BibitemShut {NoStop}%
\bibitem [{\citenamefont {Musavi}\ \emph {et~al.}(1992)\citenamefont {Musavi},
  \citenamefont {Ahmed}, \citenamefont {Chan}, \citenamefont {Faris},\ and\
  \citenamefont {Hummels}}]{musavi1992training}%
  \BibitemOpen
  \bibfield  {author} {\bibinfo {author} {\bibfnamefont {Mohamad~T}\
  \bibnamefont {Musavi}}, \bibinfo {author} {\bibfnamefont {Wahid}\
  \bibnamefont {Ahmed}}, \bibinfo {author} {\bibfnamefont {Khue~Hiang}\
  \bibnamefont {Chan}}, \bibinfo {author} {\bibfnamefont {Kathleen~B}\
  \bibnamefont {Faris}}, \ and\ \bibinfo {author} {\bibfnamefont {Donald~M}\
  \bibnamefont {Hummels}},\ }\bibfield  {title} {\enquote {\bibinfo {title} {On
  the training of radial basis function classifiers},}\ }\href@noop {}
  {\bibfield  {journal} {\bibinfo  {journal} {Neural networks}\ }\textbf
  {\bibinfo {volume} {5}},\ \bibinfo {pages} {595--603} (\bibinfo {year}
  {1992})}\BibitemShut {NoStop}%
\bibitem [{\citenamefont {Garc{\'\i}a-P{\'e}rez}\ \emph
  {et~al.}(2021)\citenamefont {Garc{\'\i}a-P{\'e}rez}, \citenamefont {Rossi},
  \citenamefont {Sokolov}, \citenamefont {Tacchino}, \citenamefont
  {Barkoutsos}, \citenamefont {Mazzola}, \citenamefont {Tavernelli},\ and\
  \citenamefont {Maniscalco}}]{garcia2021learning}%
  \BibitemOpen
  \bibfield  {author} {\bibinfo {author} {\bibfnamefont {Guillermo}\
  \bibnamefont {Garc{\'\i}a-P{\'e}rez}}, \bibinfo {author} {\bibfnamefont
  {Matteo~AC}\ \bibnamefont {Rossi}}, \bibinfo {author} {\bibfnamefont {Boris}\
  \bibnamefont {Sokolov}}, \bibinfo {author} {\bibfnamefont {Francesco}\
  \bibnamefont {Tacchino}}, \bibinfo {author} {\bibfnamefont {Panagiotis~Kl}\
  \bibnamefont {Barkoutsos}}, \bibinfo {author} {\bibfnamefont {Guglielmo}\
  \bibnamefont {Mazzola}}, \bibinfo {author} {\bibfnamefont {Ivano}\
  \bibnamefont {Tavernelli}}, \ and\ \bibinfo {author} {\bibfnamefont
  {Sabrina}\ \bibnamefont {Maniscalco}},\ }\bibfield  {title} {\enquote
  {\bibinfo {title} {Learning to measure: Adaptive informationally complete
  generalized measurements for quantum algorithms},}\ }\href@noop {} {\bibfield
   {journal} {\bibinfo  {journal} {Prx quantum}\ }\textbf {\bibinfo {volume}
  {2}},\ \bibinfo {pages} {040342} (\bibinfo {year} {2021})}\BibitemShut
  {NoStop}%
\bibitem [{\citenamefont {Huang}\ \emph {et~al.}(2020)\citenamefont {Huang},
  \citenamefont {Kueng},\ and\ \citenamefont {Preskill}}]{huang2020predicting}%
  \BibitemOpen
  \bibfield  {author} {\bibinfo {author} {\bibfnamefont {Hsin-Yuan}\
  \bibnamefont {Huang}}, \bibinfo {author} {\bibfnamefont {Richard}\
  \bibnamefont {Kueng}}, \ and\ \bibinfo {author} {\bibfnamefont {John}\
  \bibnamefont {Preskill}},\ }\bibfield  {title} {\enquote {\bibinfo {title}
  {Predicting many properties of a quantum system from very few
  measurements},}\ }\href@noop {} {\bibfield  {journal} {\bibinfo  {journal}
  {Nature Physics}\ }\textbf {\bibinfo {volume} {16}},\ \bibinfo {pages}
  {1050--1057} (\bibinfo {year} {2020})}\BibitemShut {NoStop}%
\bibitem [{\citenamefont {Romero}\ \emph {et~al.}(2017)\citenamefont {Romero},
  \citenamefont {Olson},\ and\ \citenamefont
  {Aspuru-Guzik}}]{romero2017quantum}%
  \BibitemOpen
  \bibfield  {author} {\bibinfo {author} {\bibfnamefont {Jonathan}\
  \bibnamefont {Romero}}, \bibinfo {author} {\bibfnamefont {Jonathan~P}\
  \bibnamefont {Olson}}, \ and\ \bibinfo {author} {\bibfnamefont {Alan}\
  \bibnamefont {Aspuru-Guzik}},\ }\bibfield  {title} {\enquote {\bibinfo
  {title} {Quantum autoencoders for efficient compression of quantum data},}\
  }\href {\doibase 10.1088/2058-9565/aa8072} {\bibfield  {journal} {\bibinfo
  {journal} {Quantum Sci. Technol.}\ }\textbf {\bibinfo {volume} {2}},\
  \bibinfo {pages} {045001} (\bibinfo {year} {2017})}\BibitemShut {NoStop}%
\bibitem [{\citenamefont {Deng}(2012)}]{deng2012mnist}%
  \BibitemOpen
  \bibfield  {author} {\bibinfo {author} {\bibfnamefont {Li}~\bibnamefont
  {Deng}},\ }\bibfield  {title} {\enquote {\bibinfo {title} {The mnist database
  of handwritten digit images for machine learning research [best of the
  web]},}\ }\href@noop {} {\bibfield  {journal} {\bibinfo  {journal} {IEEE
  Signal Processing Magazine}\ }\textbf {\bibinfo {volume} {29}},\ \bibinfo
  {pages} {141--142} (\bibinfo {year} {2012})}\BibitemShut {NoStop}%
\bibitem [{\citenamefont {Self}\ and\ \citenamefont
  {Haug}({\natexlab{a}})}]{self2021qml}%
  \BibitemOpen
  \bibfield  {author} {\bibinfo {author} {\bibfnamefont {Christopher}\
  \bibnamefont {Self}}\ and\ \bibinfo {author} {\bibfnamefont {Tobias}\
  \bibnamefont {Haug}},\ }\href@noop {} {\enquote {\bibinfo {title} {Code for
  large-scale quantum machine learning},}\ }\bibinfo {howpublished}
  {\url{https://github.com/chris-n-self/large-scale-qml}}
  ({\natexlab{a}})\BibitemShut {NoStop}%
\bibitem [{\citenamefont {Self}\ and\ \citenamefont
  {Haug}({\natexlab{b}})}]{self2021zenodoqml}%
  \BibitemOpen
  \bibfield  {author} {\bibinfo {author} {\bibfnamefont {Christopher}\
  \bibnamefont {Self}}\ and\ \bibinfo {author} {\bibfnamefont {Tobias}\
  \bibnamefont {Haug}},\ }\href@noop {} {\enquote {\bibinfo {title} {Data for
  large-scale quantum machine learning},}\ }\bibinfo {howpublished}
  {\url{https://doi.org/10.5281/zenodo.5211695}} ({\natexlab{b}})\BibitemShut
  {NoStop}%
\bibitem [{\citenamefont {Abraham}\ \emph {et~al.}(2019)\citenamefont {Abraham}
  \emph {et~al.}}]{Qiskit}%
  \BibitemOpen
  \bibfield  {author} {\bibinfo {author} {\bibfnamefont {H{\'e}ctor}\
  \bibnamefont {Abraham}} \emph {et~al.},\ }\href {\doibase
  10.5281/zenodo.2562110} {\enquote {\bibinfo {title} {Qiskit: An open-source
  framework for quantum computing},}\ } (\bibinfo {year} {2019})\BibitemShut
  {NoStop}%
\bibitem [{\citenamefont {Sivarajah}\ \emph {et~al.}(2020)\citenamefont
  {Sivarajah}, \citenamefont {Dilkes}, \citenamefont {Cowtan}, \citenamefont
  {Simmons}, \citenamefont {Edgington},\ and\ \citenamefont
  {Duncan}}]{sivarajah2020t}%
  \BibitemOpen
  \bibfield  {author} {\bibinfo {author} {\bibfnamefont {Seyon}\ \bibnamefont
  {Sivarajah}}, \bibinfo {author} {\bibfnamefont {Silas}\ \bibnamefont
  {Dilkes}}, \bibinfo {author} {\bibfnamefont {Alexander}\ \bibnamefont
  {Cowtan}}, \bibinfo {author} {\bibfnamefont {Will}\ \bibnamefont {Simmons}},
  \bibinfo {author} {\bibfnamefont {Alec}\ \bibnamefont {Edgington}}, \ and\
  \bibinfo {author} {\bibfnamefont {Ross}\ \bibnamefont {Duncan}},\ }\bibfield
  {title} {\enquote {\bibinfo {title} {tket: a retargetable compiler for nisq
  devices},}\ }\href@noop {} {\bibfield  {journal} {\bibinfo  {journal}
  {Quantum Science and Technology}\ }\textbf {\bibinfo {volume} {6}},\ \bibinfo
  {pages} {014003} (\bibinfo {year} {2020})}\BibitemShut {NoStop}%
\bibitem [{\citenamefont {Wallman}\ and\ \citenamefont
  {Emerson}(2016)}]{wallman2016noise}%
  \BibitemOpen
  \bibfield  {author} {\bibinfo {author} {\bibfnamefont {Joel~J}\ \bibnamefont
  {Wallman}}\ and\ \bibinfo {author} {\bibfnamefont {Joseph}\ \bibnamefont
  {Emerson}},\ }\bibfield  {title} {\enquote {\bibinfo {title} {Noise tailoring
  for scalable quantum computation via randomized compiling},}\ }\href@noop {}
  {\bibfield  {journal} {\bibinfo  {journal} {Physical Review A}\ }\textbf
  {\bibinfo {volume} {94}},\ \bibinfo {pages} {052325} (\bibinfo {year}
  {2016})}\BibitemShut {NoStop}%
\bibitem [{\citenamefont {van~den Berg}\ \emph {et~al.}(2022)\citenamefont
  {van~den Berg}, \citenamefont {Minev},\ and\ \citenamefont
  {Temme}}]{van2022model}%
  \BibitemOpen
  \bibfield  {author} {\bibinfo {author} {\bibfnamefont {Ewout}\ \bibnamefont
  {van~den Berg}}, \bibinfo {author} {\bibfnamefont {Zlatko~K}\ \bibnamefont
  {Minev}}, \ and\ \bibinfo {author} {\bibfnamefont {Kristan}\ \bibnamefont
  {Temme}},\ }\bibfield  {title} {\enquote {\bibinfo {title} {Model-free
  readout-error mitigation for quantum expectation values},}\ }\href@noop {}
  {\bibfield  {journal} {\bibinfo  {journal} {Physical Review A}\ }\textbf
  {\bibinfo {volume} {105}},\ \bibinfo {pages} {032620} (\bibinfo {year}
  {2022})}\BibitemShut {NoStop}%
\end{thebibliography}%
\clearpage
\appendix 

\begin{figure*}[htbp]
	\centering
	\subfigimg[width=0.99\textwidth]{}{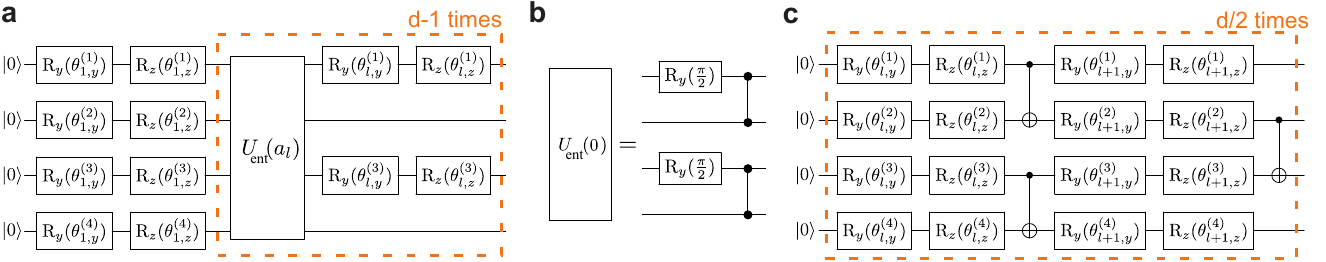}
	\caption{\idg{a} The NPQC for $N$ qubits and $d$ layers, which is a hardware efficient PQC composed of single qubit rotations and CPHASE gates. For the reference parameter $\boldsymbol{\theta}_\text{r}$, the QFIM is the identity matrix.
	\idg{b} Example of the entangling layer for the NPQC, which is composed of $N/2$ non-overlapping CPHASE gates and $y$ rotations by $\pi/2$.
	\idg{c} YZ-CX PQC, which is a hardware efficient circuit consisting of single qubit rotations and CNOT gates arranged in an alternating fashion in even and odd layers $l$.
	}
	\label{fig:circuits}
\end{figure*}

\section{Parameterized quantum circuits}\label{sec:PQC}

\revB{Here, we describe the two types of PQCs used in the main text. The PQCs are composed of $N$ qubits and $d$ layers of unitaries. The parameters of the PQC are given by the $M$-dimensional parameter vector $\boldsymbol{\theta}\in\mathbb{R}^M$.
Each layer is described by unitary $U_l(\boldsymbol{\theta}_l)$ with the parameter vector of each layer $\boldsymbol{\theta}_l$ with $\boldsymbol{\theta}=\{\boldsymbol{\theta}_1,\dots,\boldsymbol{\theta}_d\}$. The total PQC is given by
$U(\boldsymbol{\theta})\ket{0}=\prod_{l=1}^d U_l(\boldsymbol{\theta}_l)\ket{0}^{\otimes N}$. Each layer unitary $U_l(\boldsymbol{\theta}^l)$ is is composed of parameterized single qubit rotations and an unparameterized entangling gate.
For each layer, we denote each parameter entry by $\theta_{l,\alpha}^{(k)}$, where $l$ denotes the layer, $\alpha\in\{x,y,z\}$ the type of rotation and $k$ the qubit number. Note this notation is different from the main text.}

In Fig.\ref{fig:circuits}a, we show the first circuit we use, which we call the NPQC.
The first layer is composed of $2N$ single qubit rotations around the $y$ and $z$ axis for each qubit $n$ with $U_1=\prod_{k=1}^N R_z^{(k)}(\theta_{1,z}^{(k)})R_y^{(k)}(\theta_{1,y}^{(k)})$. Here, $R_\alpha^{(k)}(\theta)=\exp(-i\frac{\theta}{2}\sigma^\alpha_n)$, $\alpha\in\{x,y,z\}$ and $\sigma^x_n$, $\sigma^y_n$, $\sigma^z_n$ are the Pauli matrices acting on qubit $k$.
Each additional layer $l>1$ is a product of two qubit entangling gates and $N$ parameterized single qubit rotations defined as
$U_l(a_l)=\prod_{k=1}^{N/2}[R_z^{(2k-1)}(\theta_{l,z}^{(2k-1)})R_y^{(2k-1)}(\theta_{l,y}^{(2k-1)})]U_\text{ent}(a_l)$, where $U_\text{ent}(a_l)=\prod_{k=1}^{N/2}\text{CPHASE}(2k-1,2k+2a_l)R_y^{(2k-1)}(\pi/2)$ and $\text{CPHASE}(n,m)$ is the controlled $\sigma^z$ gate for qubit index $n$, $m$, where indices larger than $N$ are taken modulo. The entangling layer $U_l(0)$ is shown as example in Fig.\ref{fig:circuits}b.
The shift factor $a_l\in\{0,1,\dots,N/2-1\}$ for layer $l$ is given by the recursive rule shown in the following.
Initialise a set $A=\{0,1,\dots,N/2-1\}$ and $s=1$. In each iteration, pick and remove one element $r$ from $A$. 
Then set $a_s=r$ and $a_{s+q}=a_{q}$ for $q=\{1,\dots,s-1\}$. As the last step, we set $s=2s$. We repeat this procedure until no elements are left in $A$ or a target depth $d$ is reached. 
One can have maximally $d_\text{max}=2^{N/2}$ layers with in total $M=N(d+1)$ parameters.
The NPQC has a QFIM $\mathcal{F}(\boldsymbol{\theta_\text{r}})=I$, $I$ being the identity matrix, for the reference parameter $\boldsymbol{\theta}_\text{r}$ given by
\begin{equation}\label{eq:ref_params}
\mathcal{F}(\boldsymbol{\theta}_\text{r})=I\hspace{0.2cm}\text{for}\hspace{0.2cm} \theta_{\text{r},l,y}^{(k)}=\pi/2\,,\hspace{0.2cm}\theta_{\text{r},l,z}^{(k)}=0 \,,
\end{equation}
\revB{where $\theta_{\text{r},l,z}^{(k)}$ is the reference parameter for layer $L$, qubit $k$ and rotation around $z$-axis.}
Close to this reference parameter, the QFIM remains approximately close to being an identity matrix.
When implementing the NPQC for the IBM quantum computer, we choose the sift factor $a_l$ such that only nearest-neighbor CPHASE gates arranged in a chain appear. To match the connectivity of the IBM quantum computer, we removed one entangling gate and its corresponding single qubit rotations which require connection between the first and the last qubit of the chain.

The second type of PQC used is shown in Fig.\ref{fig:circuits}c, which we call YZ-CX. It consists of $d$ layers of parameterized single qubit $y$ and $z$ rotations, followed by CNOT gates. The CNOT gates arranged in a one-dimensional chain, acting on neighboring qubits. Every layer $l$, the CNOT gates are shifted by one qubit. Redundant single qubit rotations that are left over at the edges of the chain are removed.

\section{Methods to measure quantum kernels}\label{sec:meas_kernel}
\begin{figure}[htbp]
	\centering
	\subfigimg[width=0.37\textwidth]{}{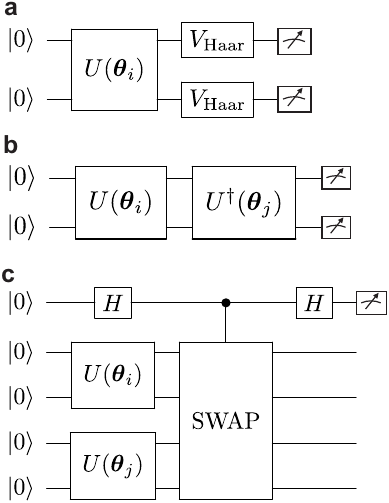}
	\caption{Quantum circuits to measure kernel $K(\boldsymbol{\theta}_i,\boldsymbol{\theta}_j)$ of $L$ quantum states $U(\boldsymbol{\theta}_i)\ket{0}$.
	\idg{a}  Randomized measurement scheme. Prepare state $\ket{\psi(\boldsymbol{\theta}_i)}$, rotate into randomized basis given by single qubit Haar random unitaries $V_\text{Haar}$ and measure in computational basis. By post-processing sampled states one can extract the kernel. The number of measurements scales linearly with $L$.
	\idg{b} Inversion test. Prepare $U^\dagger(\boldsymbol{\theta}_j)U(\boldsymbol{\theta}_i)\ket{0}$ and measure probability of zero state $\ket{0}$. The number of measurements scale with $L^2$.
	\idg{c} Swap test. Prepare $U(\boldsymbol{\theta}_j)\otimes U(\boldsymbol{\theta}_i)\ket{0}$ on twice the number of qubits and perform controlled SWAP gate with an ancilla. Kernel is determined by measuring ancilla only. The number of measurements scale with $L^2$.
	}
	\label{fig:meas_circuit}
\end{figure}

In Fig.\ref{fig:meas_circuit}, we explain the different methods to measure kernels of $L$ quantum states. In this paper, we use the randomized measurements method shown in Fig.\ref{fig:meas_circuit}a. The number of required measurements to measure all possible pairs of kernels scales linearly with dataset size $L$. 

The inversion test is shown in Fig.\ref{fig:meas_circuit}b. To measure the kernel between two quantum states, it uses the unitary of the first state combined the with inverse unitary of the second state. Then, the kernel is given by the probability of measuring the zero state. Here, the number of measurements scales with the square $L^2$ of the dataset size.

The swap test is shown in Fig.\ref{fig:meas_circuit}c. It prepares both states for the kernel, requiring two times the amount of qubits as with the other tests. Then, a controlled SWAP gate is applied, with the control being on an ancilla qubit. Then, the kernel is given by the measurement of the ancilla.
As with the inversion test, the number of required measurements scales with the square $L^2$ of the dataset size. Further, the controlled SWAP gate can require substantial quantum resources.

\section{Experimental kernel of YZ-CX PQC}\label{sec:exp_kernel}
\begin{figure}[htbp]
	\centering	
	\subfigimg[width=0.35\textwidth]{}{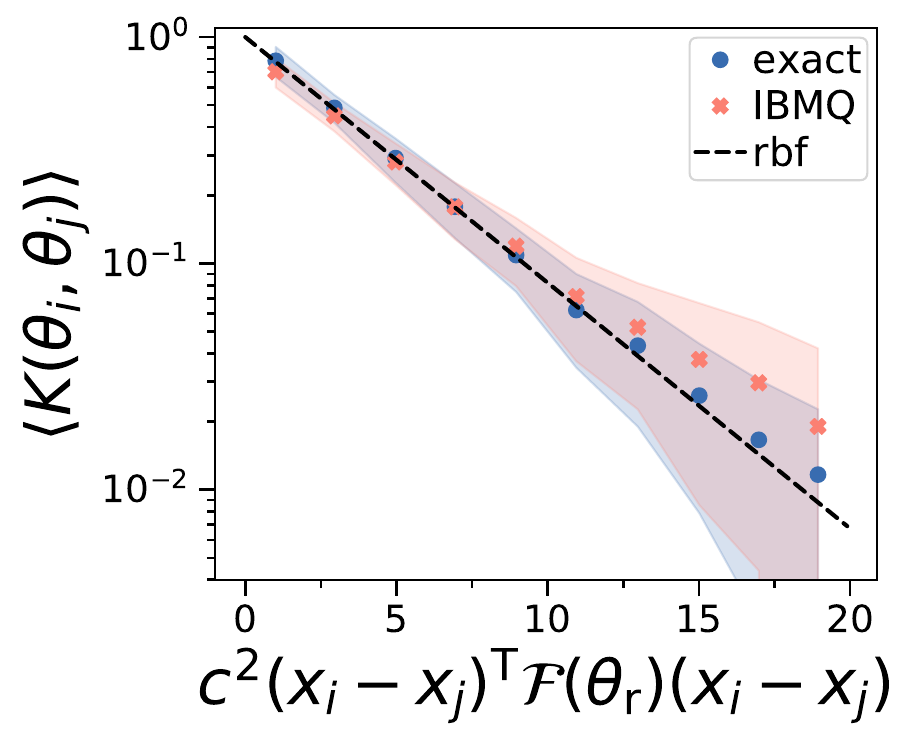} 
	\caption{ Experimental kernel $K(\boldsymbol{\theta}_i,\boldsymbol{\theta}_j)$ as function of distance between the feature vectors $\boldsymbol{x}_i$, $\boldsymbol{x}_j$. We encode randomly chosen feature vectors via $\boldsymbol{\theta}_i=\boldsymbol{\theta}_\text{r}+c\boldsymbol{x}_i$ in the YZ-CX PQC, where $\boldsymbol{\theta}_\text{r}\in[0,2\pi]$ is a randomly chosen reference parameter. The QFIM $\mathcal{F}(\boldsymbol{\theta}_\text{r})$ is calculated numerically. The quantum kernel generated by theory (blue dots) and via experimental results with IBM quantum computer (orange crosses) follows approximately a weighted radial basis function kernel $K(\boldsymbol{\theta}_i,\boldsymbol{\theta}_j)=\exp[-\frac{c^2}{4}(\boldsymbol{x}_i-\boldsymbol{x}_j)^{\text{T}}\mathcal{F}(\boldsymbol{\theta}_\text{r})(\boldsymbol{x}_i-\boldsymbol{x}_j)]$ (black line). Shaded area is standard deviation of the kernel. The YZ-CX PQC has $N=8$ qubits, $M=72$ features and $d=5$ layers. Experimental results from \emph{ibmq\_guadalupe} were performed with $r=50$ randomized measurement settings, $s=8192$ measurement samples and error mitigation. 
	}
	\label{fig:yzkernel}
\end{figure}

We provide further data on the experimental quantum kernel measured on the IBM quantum computer. We measure the kernel using randomized measurements for randomly chosen feature vectors.
In Fig.\ref{fig:yzkernel}, we show experimental data of the kernel for the YZ-CX PQC using \emph{ibmq\_guadalupe}. We find that the experimental data and numerical simulations match well.

\section{Measurement cost}\label{sec:cost}
\begin{figure}[htbp]
	\centering	
	\subfigimg[width=0.35\textwidth]{}{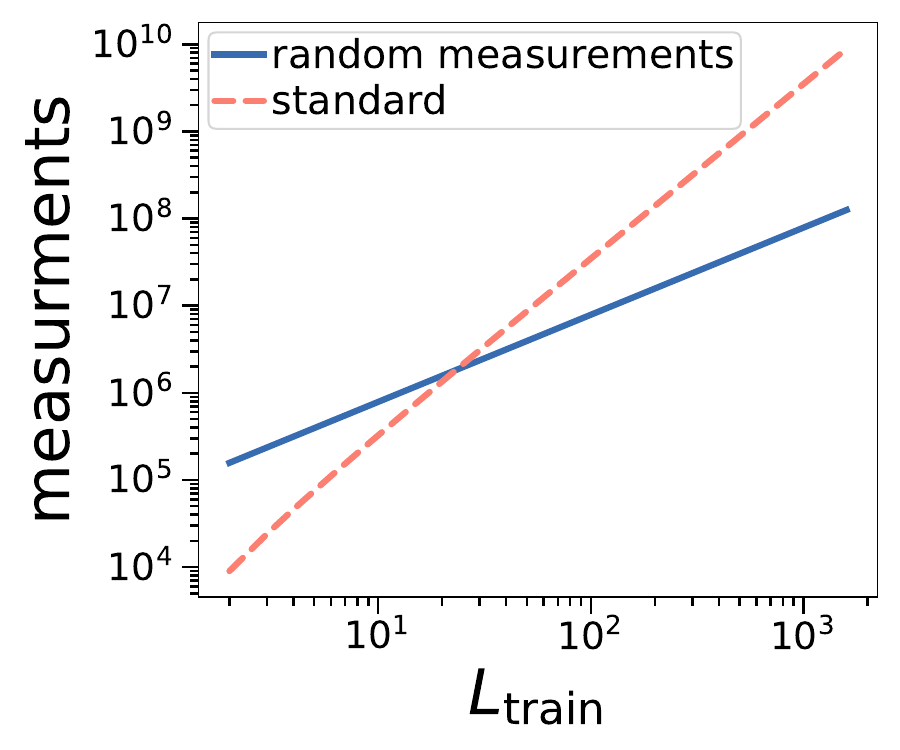} 
	\caption{\revB{Number of measurements for randomized and standard measurement methods as function of training data $L_\text{train}$ measured on the quantum computer. We set test data to $L_\text{test}=L_\text{train}/5$ and use $s=8192$, $r=8$ and $s_\text{c}=5000$.}
	}
	\label{fig:cost}
\end{figure}

\revB{Here, we compare the measurement cost when learning from our dataset for varying number of training data used.
For randomized measurements, the number of measurements is given by $N_\text{meas}^\text{random}=s(L_\text{train}+L_\text{test})r$. For conventional methods such as SWAP or inversion test, we have
$N_\text{meas}^\text{inv}=s_\text{c}L_\text{train}(L_\text{train}-1)/2+s_\text{c}L_\text{train}L_\text{test}$. We now assume that $L_\text{test}=L_\text{train}/5$. We assume $s=8192$ and $r=8$ with the same values as used for experiment of $N=9$ qubits in the main text. For the conventional approach, we choose $s_\text{c}=5000$ as used in~\cite{peters2021machine} for an experiment with comparable feature vector size.  The measurement cost is plotted in Fig.\ref{fig:cost}, where we find that randomized measurements is advantageous with $N_\text{meas}^\text{random}<N_\text{meas}^\text{inv}$ for $L_\text{train}>21$.}

\section{IBM Quantum implementation details}\label{sec:ibm_details}

Our PQC circuits are constructed as parameterised circuits with Qiskit~\cite{Qiskit}. These parameterised circuits are first transpiled then bound for each data point and randomized measurement unitary, ensuring that all circuits submitted have the same structure and use the same set of device qubits. Transpiling is handled by the pytket python package~\cite{sivarajah2020t} using \emph{rebase}, \emph{placement} and  \emph{routing} passes with no additional optimisations (IBMQ default passes with optimisation level 0).

The \emph{ibmq\_guadalupe}~\cite{ibmq-devices} results presented in Fig.~\ref{fig:kernel} and Fig.~\ref{fig:accuracy} were collected between 22nd July 2021 and 30th July 2021. The \emph{ibmq\_toronto}~\cite{ibmq-devices} results presented in Fig.~\ref{fig:accuracy} were collected between 23rd July 2021 and 9th August 2021. 
Fig.~\ref{fig:kernel} required the execution of $100 \times 50 = 5000$ circuits and Fig.~\ref{fig:accuracy} involved $1790 \times 8 = 14,320$ circuits, each with 8192 measurement shots. For comparison, applying the inversion test to the same handwritten digit dataset used for Fig.~\ref{fig:accuracy} would have required the execution of $1790 \times 1790 \approx 3.2 \times 10^6$ circuits.
Circuits were executed on IBM quantum devices using the circuit queue API. Job submissions were batched in such a way that all measurement circuits for a data point were submitted and executed together. 

Beyond the error mitigation procedure described in the main text we carry out no further error mitigation. \revB{In particular, we find that within our experiments readout error mitigation does not yield any significant advantages. We attribute this two possible origins. Our randomized measurement scheme applies random unitaries, which effectively twirl the noise into a form which is easier to mitigate and has been shown to substantially reduce errors~\cite{wallman2016noise}. Further, data collected from application of random Pauli operators subject to the same noise has been shown to efficiently correct read-out errors~\cite{van2022model}. Similar to this approach, it is possible that our error mitigation scheme also corrects read-out errors at the same time. 
%We attribute this to the fact that to estimate the kernel using randomized measurements, we multiply probability distributions sampled from the quantum computer, which are affected by the same error. This could lead to an effective canceling out of the readout errors. 
}

\begin{figure}[htbp]
	\centering	
	\subfigimg[width=0.24\textwidth]{a}{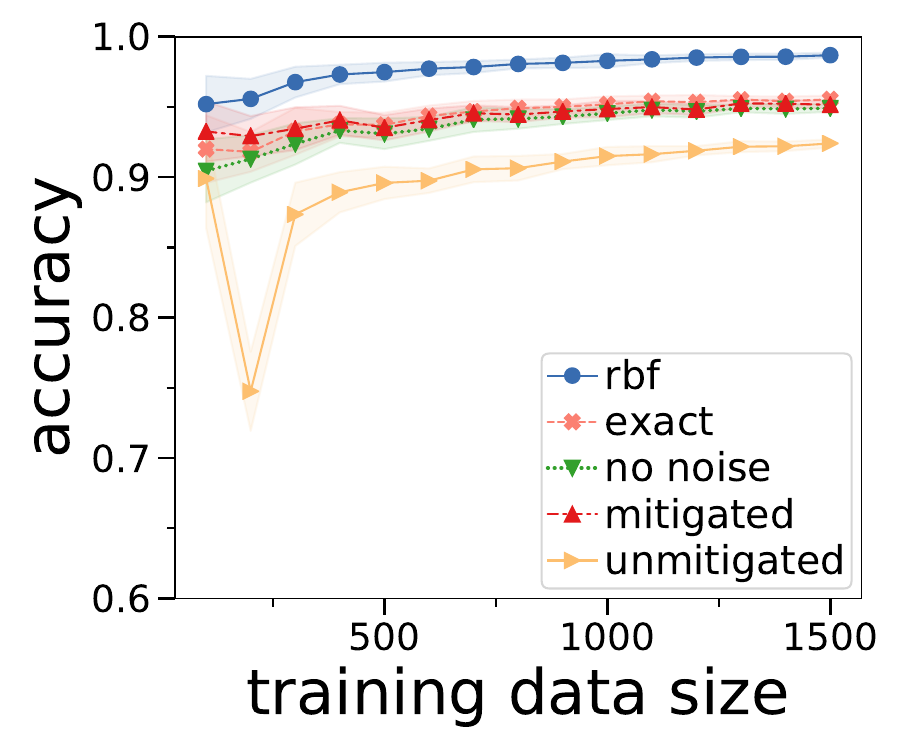}\hfill
	\subfigimg[width=0.24\textwidth]{b}{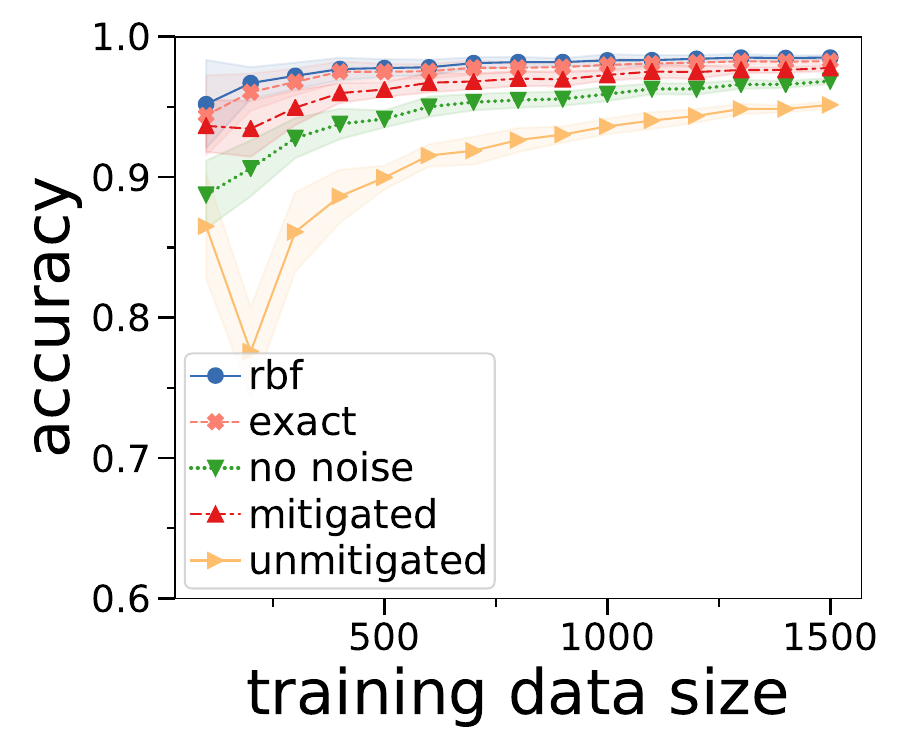}\\	\subfigimg[width=0.24\textwidth]{c}{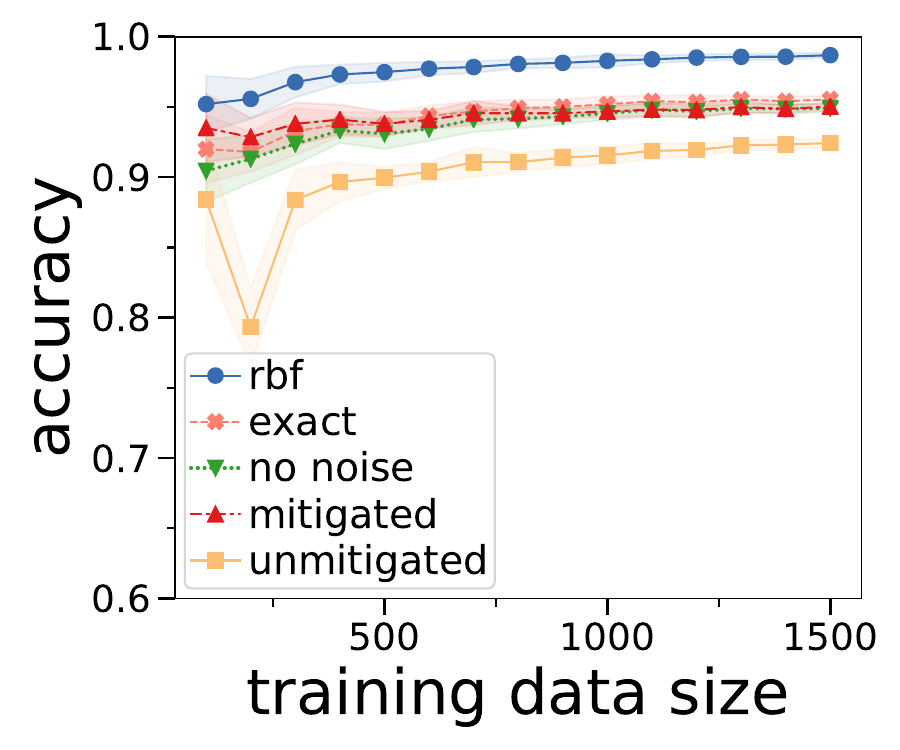}\hfill
	\subfigimg[width=0.24\textwidth]{d}{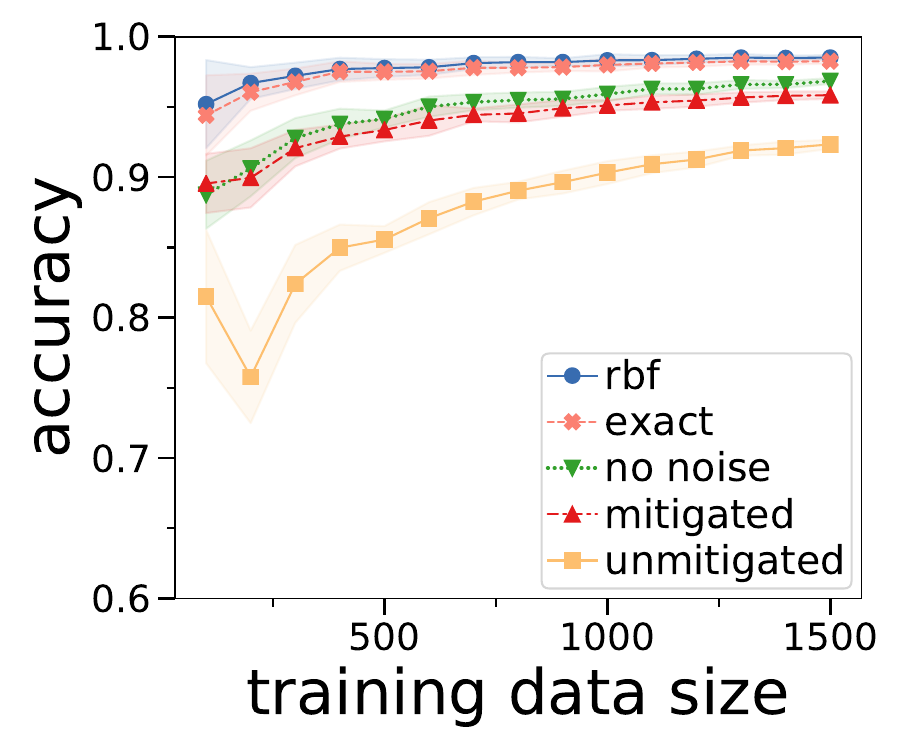}\\
	\subfigimg[width=0.24\textwidth]{e}{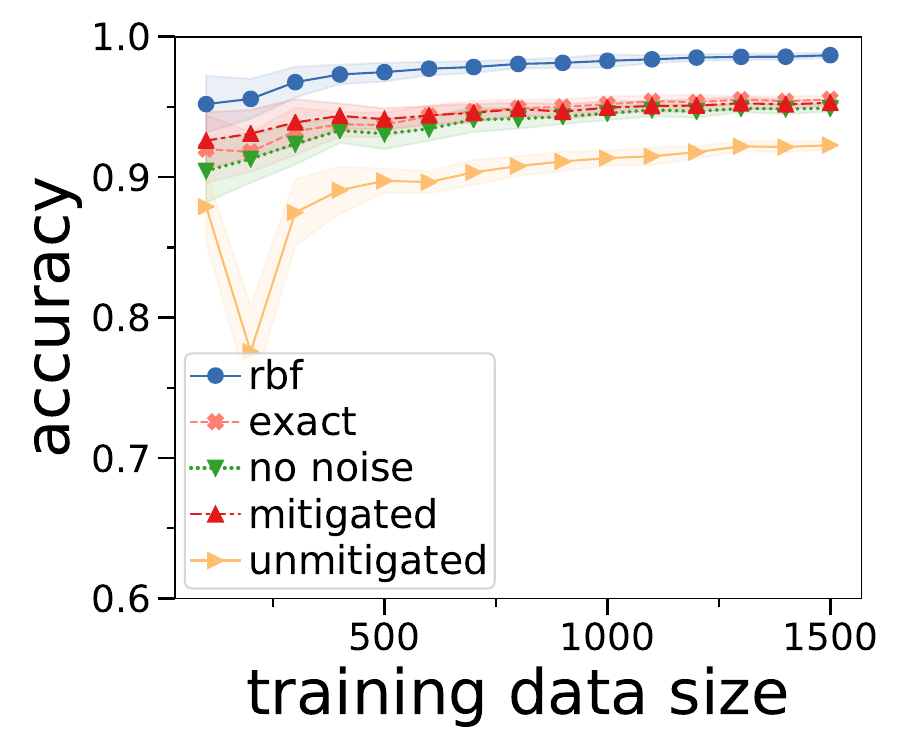}\hfill
	\subfigimg[width=0.24\textwidth]{f}{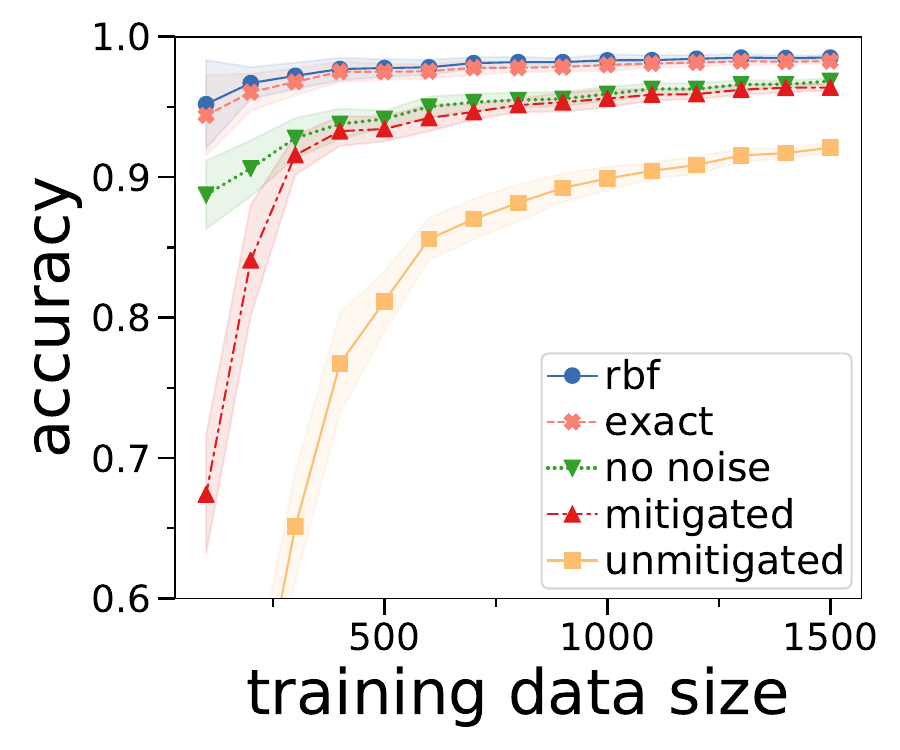}
	\caption{Accuracy of classifying training data as function of the size of the training data. The shaded area is the standard deviation of the accuracy. We use same data as in the main text. We compare the radial basis function kernel (blue dots), exact quantum kernel (orange), noiseless simulation of randomized measurements (green), kernel of IBM quantum computer with randomized measurements using error mitigation (red) and no error mitigation (yellow). 
	\idg{a,c,e} YZ-CX PQC with $M=64$ features  and \idg{b,d,f} NPQC with $M=36$ features. We use \idg{a,b} \emph{ibmq\_guadalupe}, \idg{c,d} \emph{ibmq\_toronto} and \idg{e,f} equally distributing measurements between aforementioned quantum computers. 
	We have $s=8192$ measurement samples, $N=8$ qubits and $r=8$ randomized measurement settings. The training data is randomly drawn from the full dataset, which is repeated 20 times for each training data size.
	}
	\label{fig:accuracy_training}
\end{figure}

\section{Training accuracy}\label{sec:training}
In Fig.\ref{fig:accuracy_training}, we plot the accuracy of classifying the training data with the SVM for the YZ-CX PQC and NPQC. We show the accuracy for processing on \emph{ibmq\_guadalupe}, \emph{ibmq\_toronto} and distributing the dataset between both quantum computers. The accuracy is defined as the percentage of training data that is correctly identified. We find that error mitigation substantially increases the accuracy in all cases.

\section{Confusion matrix}\label{sec:confusion}
\begin{figure*}[htbp]
	\centering	
	\subfigimg[width=0.24\textwidth]{a}{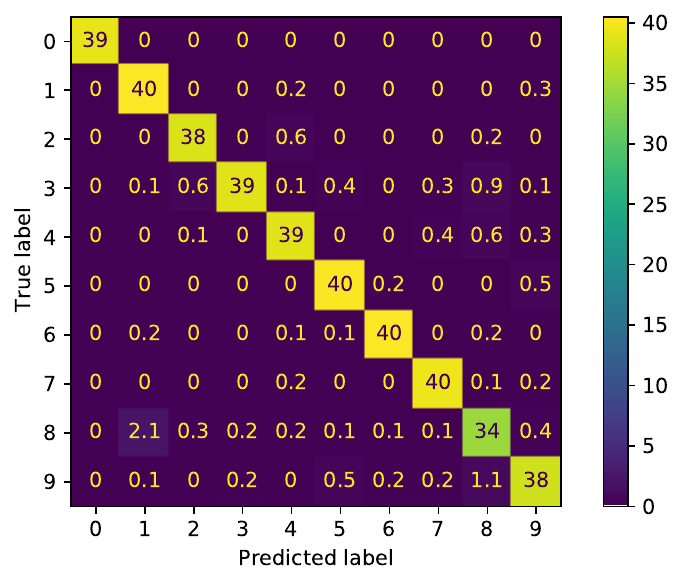}\hfill
	\subfigimg[width=0.24\textwidth]{b}{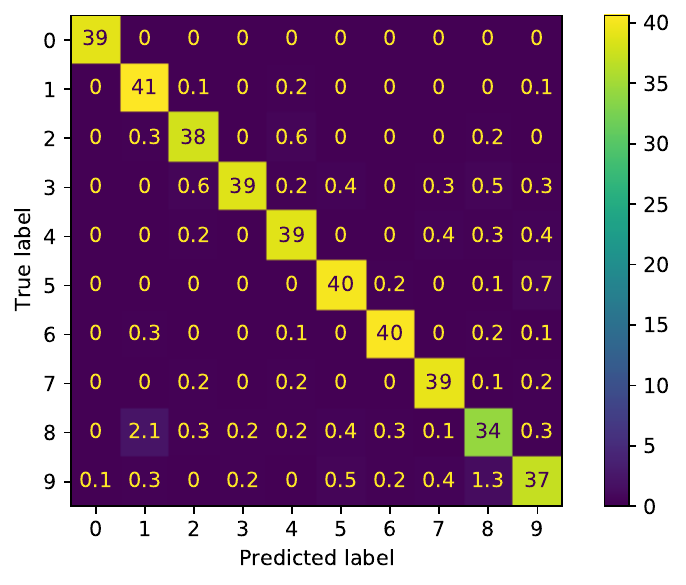}\hfill
	\subfigimg[width=0.24\textwidth]{c}{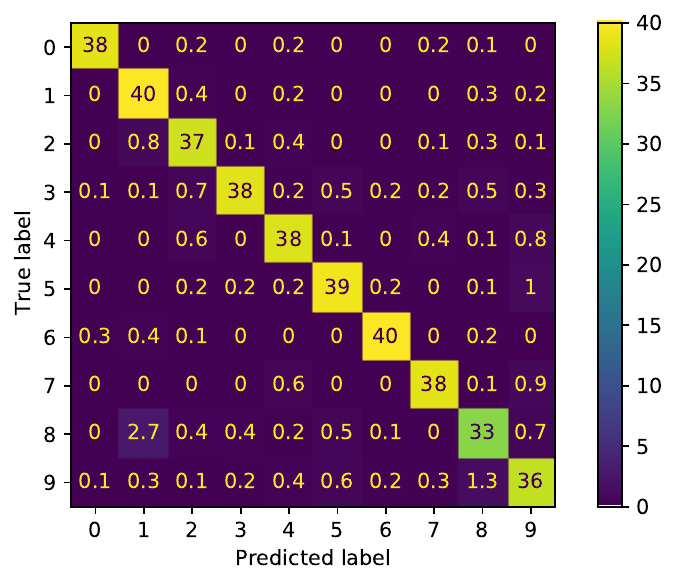}\hfill
	\subfigimg[width=0.24\textwidth]{d}{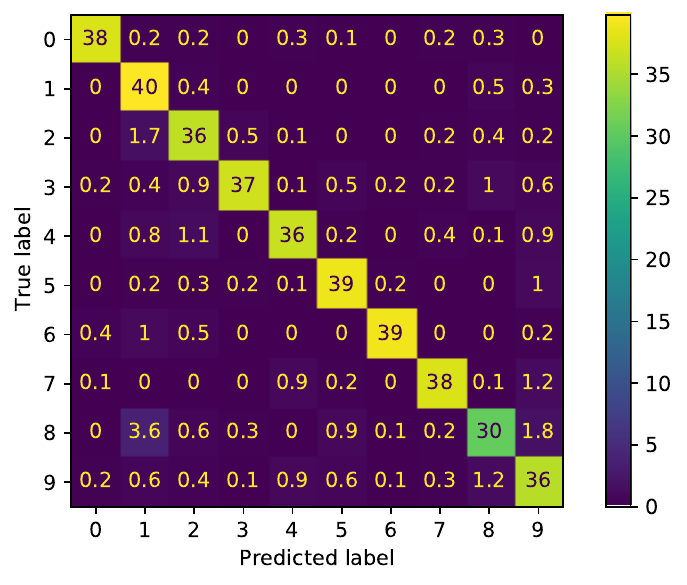}
	\caption{Confusion matrix for the NPQC for \idg{a} radial basis function kernel \idg{b} exact quantum kernel \idg{c} mitigated IBM results with \emph{ibmq\_guadalupe} and \idg{d} unmitigated IBM results. We use 1300 training data and 200 test data, where we average the confusion matrix over 100 randomly sampled instances of the data.
	}
	\label{fig:confusion_NPQC}
\end{figure*}

\begin{figure*}[htbp]
	\centering	
	\subfigimg[width=0.24\textwidth]{a}{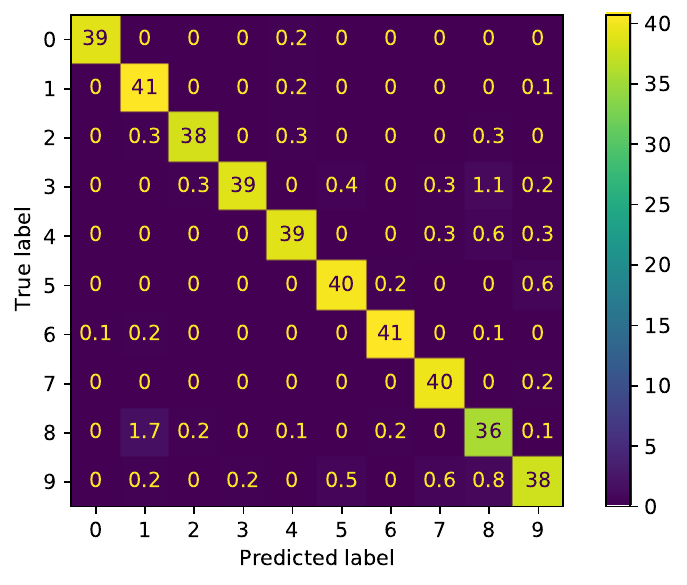}\hfill
	\subfigimg[width=0.24\textwidth]{b}{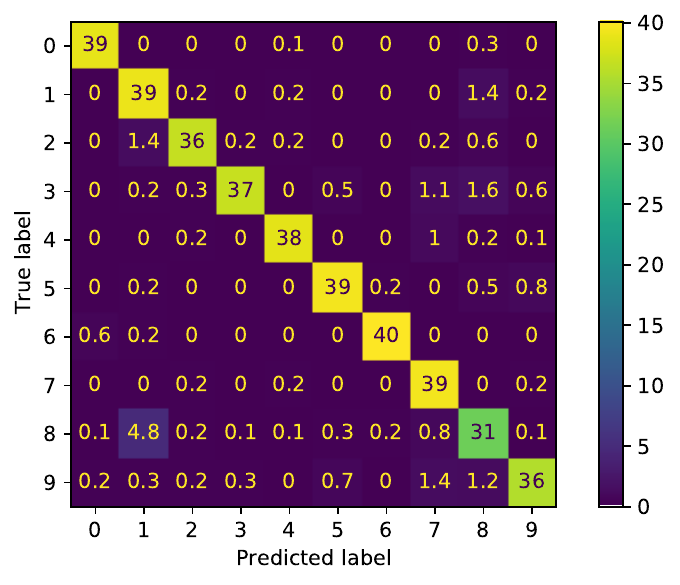}\hfill
	\subfigimg[width=0.24\textwidth]{c}{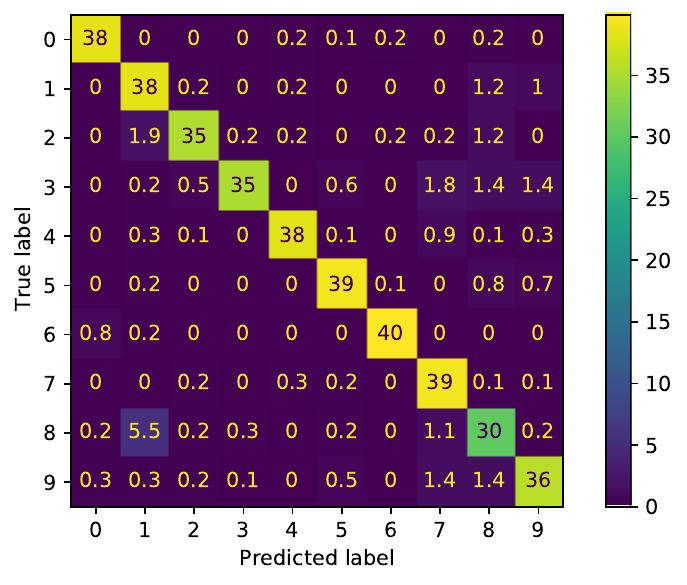}\hfill
	\subfigimg[width=0.24\textwidth]{d}{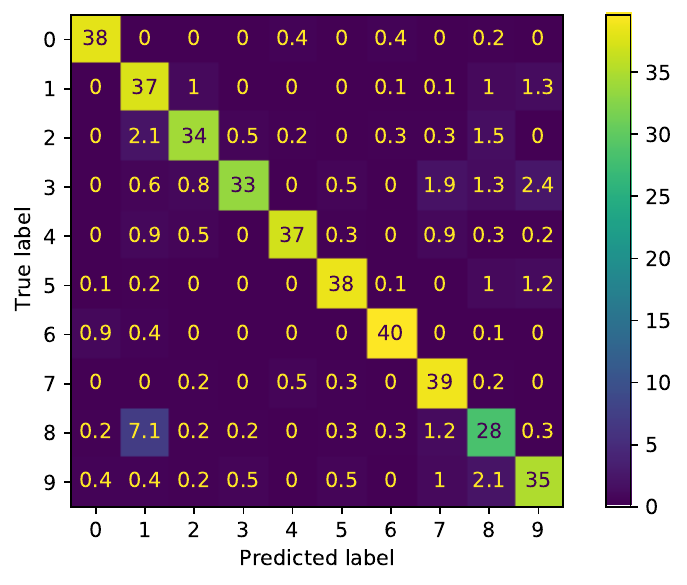}
	\caption{Confusion matrix for the YZ-CX PQC for \idg{a} radial basis function kernel \idg{b} exact quantum kernel \idg{c} mitigated IBM results with \emph{ibmq\_guadalupe} and \idg{d} unmitigated IBM results. We use 1300 training data and 200 test data, where we average the confusion matrix over 100 randomly sampled instances of the data.
	}
	\label{fig:confusion_YZ}
\end{figure*}

We now show the confusion matrices for the test data. The confusion matrix shows what label is predicted by the SVM in respect to its true label of the test data. The diagonal are the correctly classified digits, whereas the off-diagonals show the number of times a digit was miss-classified.
In Fig.\ref{fig:confusion_NPQC}, we show the confusion matrix for the NPQC.
and in Fig.\ref{fig:confusion_YZ} we show the confusion matrix for the YZ-CX PQC. 
We find that the actual digit 8 is often predicted to be the digit 1. Then, likely confusions are that digit 3 is assumed to be 8 and digit 9 is assumed to be 8. We find these confusions consistently in all kernels. 
While for the NPQC, radial basis function kernel and quantum kernel give nearly the same confusion matrix, we find substantial differences for the YZ-CX PQC. The reason is that while NPQC is an approximate isotropic radial basis function kernel, the YZ-CX PQC is an approximate radial basis function kernel with a weight matrix given by the QFIM. The weight matrix of the YZ-CX seems to reduce the accuracy of the trained SVM.

\section{Product state as analytic radial basis function kernel}\label{sec:product_state}
As an analytic example, we show that product states form an exact radial basis function kernel. We use the following $N$ qubit quantum state 
\begin{equation}
\ket{\psi(\boldsymbol{\theta})}=\bigotimes_{n=1}^N(\cos(\frac{\theta_n}{2})\ket{0}+\sin(\frac{\theta_n}{2})\ket{1})\,.
\end{equation}
The QFIM is given by $\mathcal{F}(\boldsymbol{\theta})=I$, where $I$ is the $M$-dimensional identity matrix and $\mathcal{F}_ {ij}=4[\braket{\partial_i \psi}{\partial_j \psi}-\braket{\partial_i \psi}{\psi}\braket{\psi}{\partial_j \psi}]$.
The kernel of two states parameterized by $\boldsymbol{\theta}$, $\boldsymbol{\theta}'$ is given by
\begin{equation}
K(\boldsymbol{\theta},\boldsymbol{\theta}')=\abs{\braket{\psi(\boldsymbol{\theta})}{\psi(\boldsymbol{\theta}')}}^2=\prod_{n=1}^N(1+\frac{1}{2}\cos(\Delta\theta_n))
\end{equation}
where we define $\Delta\boldsymbol{\theta}=\boldsymbol{\theta}-\boldsymbol{\theta}'$ as the difference between the two parameter sets. We now assume $\abs{\Delta\theta_n}\ll 1$ and that all the differences of the parameters are equal $\Delta\theta_1=\dots=\Delta\theta_N$. We then find in the limit of many qubits $N$
\begin{equation}\label{eq:expfidelity}
K(\boldsymbol{\theta},\boldsymbol{\theta}')\approx\prod_{n=1}^N(1-\frac{1}{4}\Delta\theta_n^2)\xrightarrow[N \to \infty]{}\text{exp}(-\frac{1}{4}\sum_{n=1}^N\Delta\theta_n^2)\,,
\end{equation}
which gives us the radial basis function kernel.
%\newpage

\end{document}